\documentclass[a4paper,fleqn]{cas-sc}

\usepackage[numbers]{natbib}

\def\tsc#1{\csdef{#1}{\textsc{\lowercase{#1}}\xspace}}
\tsc{WGM}
\tsc{QE}
\tsc{EP}
\tsc{PMS}
\tsc{BEC}
\tsc{DE}

\begin{document}

\let\WriteBookmarks\relax

\title [mode = title]{Identifiability Limits in Ultrasonic Microstructure Characterisation: A Canonical and Stochastic Framework}                      



\author[1]{Wei Yi Yeoh}[type=editor]
\cormark[1]
\ead{yeohwy1@a-star.edu.sg}


\affiliation[1]{organization={Advanced Remanufacturing \& Technology Centre (ARTC), Agency for Science, Technology and Research (A*STAR)},
                addressline={3 Cleantech Loop, \#01/01 CleanTech Two}, 
                postcode={637143}, 
                country={Republic of Singapore}}
\cortext[cor1]{Corresponding author}


\begin{abstract}
Ultrasound for microstructure characterisation is increasingly studied and is often assessed through inversion performance. However, the framework is fundamentally constrained by the information content available in the measured response. Hence, this work examines identifiability directly by analysing the geometry of the forward operator in both a canonical pulse-echo model and a stochastic surrogate microstructure. For the canonical model, a closed-form sensitivity analysis reveals information limits arising from parameter coupling, dimensional restriction, and interface-driven saturation. For the surrogate microstructures represented by Gaussian random fields, the forward map from correlation length $D$ and texture-coherence parameter $T$ to the attenuation and velocity observables remains structurally full rank. However, the sensitivity geometry is strongly anisotropic, with uneven parameter influence across the observable space. When intrinsic microstructural variability is incorporated, practical identifiability is further reduced. A variance-weighted Fisher framework shows that recoverability is governed by the balance between sensitivity magnitude and stochastic variability, rather than by structural rank alone. Inversion results confirm this behaviour: single observables produce elongated and weakly constrained objective landscapes, whereas combined observables improve conditioning through complementary sensitivities. These results show that, within the feature-level framework considered here, identifiability limits are governed primarily by forward-map structure and intrinsic variability, with direct implications for observable selection and measurement design.
\end{abstract}




\maketitle
\thispagestyle{empty}

\section{Introduction}

Ultrasound is increasingly used for non-destructive evaluation (NDE) and the characterisation of material microstructure. The elastodynamic principle couples the effects of wave propagation and scattering to variations in material properties, enabling the inference of microstructure information based on ultrasound response \cite{Rose1999a}. This relationship provides the basis for ultrasonic microstructure characterisation methods that span from analytical scattering models to numerical simulations and experimental inversion approaches.

The theoretical framework for ultrasonic scattering in polycrystalline materials was established through the unified scattering theory that relates attenuation and velocity dispersion to elastic heterogeneity \cite{Stanke1984a,Weaver1990}. The general formulation was later adapted to evaluate backscatter grain noise \cite{Margetan1993a,Margetan1994}. Since then, scattering models have been extended to include more realistic microstructural descriptors, such as second-order scattering effects \cite{VanPamel2018a}, elongated grain morphologies \cite{Li2014a,liu2022shape}, grain size distributions \cite{renaud2021multiparameter,Sheng2026BroadGrainSize}, anisotropic texture \cite{victoriagiraldo2026texture}, multi-phase microstructures \cite{Lobkis2012b}, and wave modes \cite{li2024bs}. These studies show that ultrasonic observables are sensitive to microstructural features, with some inherent coupling effects between physical mechanisms.

In parallel, numerical modelling has become an important framework to validate the scattering theories \cite{VanPamel2017a,dorval2025numerical} and to explore regimes where analytical assumptions become restrictive. Finite element (FE) simulations have been used to study wave propagation and scattering in two- and three-dimensional polycrystalline media \cite{Bai2020,victoriagiraldo2025ultrasonic} that contain explicit grain morphologies generated through tessellation methods \cite{Quey2011}. Apart from the addition of complex features such as elongated grains \cite{Arguelles2016b,Ming2020}, highly anisotropic and scattering properties \cite{Huang2021a,Gong2026}, textured heterogeneities \cite{Li2015c,Yeoh2023}, and multi-phase material \cite{Liu2025_Duplex}, recent work also used generated synthetic microstructures derived from experimental microstructure data to improve the representation of simulated media \cite{grabec2022synthetic,Yeoh2025}. This modelling framework has made it possible to examine the influence of microstructure descriptors on ultrasonic responses that otherwise would be difficult to isolate experimentally.

Experimental approaches have been progressively adopted to validate the analytical and numerical models and to infer microstructural descriptors directly from ultrasonic measurements. Attenuation-based approaches are used to estimate grain sizes in metallic alloys \cite{Liu2021a,ruiz2024ultrasonic}. Backscatter-based approaches are used to characterise spatially varying microstructures \cite{liu2025ultrasonic} and anisotropic heterogeneity \cite{li2025backscatter,vela2026characterization}. Velocity-based approaches are used to evaluate elastic stiffness \cite{Lan2014a}, crystal orientations \cite{dryburgh2020velmap,Sun2024VelocityMap}, and bulk texture \cite{Bo2018}. Recent studies have increasingly used phased arrays to improve backscatter-based microstructure characterisation, where beamforming enables spatially resolved and directionally selective interrogation of scattering signatures \cite{liu2023spatial,Wang2025,liu2025combined}. Beyond conventional measurements, alternative ultrasonic modalities have also been explored for microstructure characterisation, including resonant ultrasound spectroscopy for inferring effective elastic properties \cite{Lan2018}, grazing-incidence ultrasonic microscopy for grain-scale surface interrogation \cite{kalkowski2021grazing}, and coda-wave method that exploits multiply scattered late-time signals to estimate grain-sizes \cite{He2024CodaWaveGrainSize}.

In recent years, machine-learning approaches have become increasingly popular in ultrasound. Particularly for materials characterisation, data-driven networks are used to identify grain-size classes from backscattered data \cite{Viana2024} and to infer grain-size distributions from non-linear wave responses \cite{liu2022autonomous}. A similar physics-guided simulation approach enabled the acceleration of anisotropic weld properties reconstruction from array data \cite{singh2022deep} and crystallographic orientation mapping from surface wave measurements \cite{Patel2025developing}. Physics-informed networks have also been explored for inferring spatially varying elastic properties from wavefield data and demonstrated potential for solving ill-posed inverse problems \cite{shukla2022physics}. While these studies demonstrate the improved analysis and acceleration with the use of machine-learning networks, they do not eliminate the need to understand whether the measured wavefields contain sufficient independent information about the target descriptors.

These developments demonstrate the increasing maturity of ultrasonic microstructure characterisation. However, they also highlight a central challenge: in practical engineering materials, multiple descriptors often influence the same measured wavefield in coupled and anisotropic ways. Recent studies have begun to address this complexity through multi-parameter analysis \cite{Rokhlin2021a,guo2026decoupling}. Nevertheless, it is often unclear whether the ambiguities in the inferred properties arise from limitations of the inversion method or from intrinsic non-uniqueness in the forward mapping. This motivates an identifiability-based analysis of ultrasonic microstructure characterisation. Rather than asking only whether an inversion or learning method can reproduce a target descriptor, the central question is whether the chosen ultrasonic observables contain sufficient independent and stable information to recover that descriptor in the first place.

In this work, a unified identifiability study is presented to quantify the intrinsic information content of ultrasonic measurements through the analysis of a feature-level forward map \cite{Bellman1970,Cobelli1980,Guillaume2019}. A minimal canonical model is first used to isolate the fundamental mechanisms governing sensitivity and parameter coupling, before extending the same analysis to stochastic surrogate microstructures that capture distributed scattering and variability. Across both settings, identifiability is characterised through singular-value structure and a variance-weighted Fisher framework. The results show that the main challenge to identifiability is not only dimensional restriction, but also imbalance in sensitivity across parameter directions. In the surrogate setting, the forward map remains structurally full rank, yet certain directions are intrinsically weak and become further suppressed when stochastic variability is taken into account. This leads to a loss of practical identifiability, showing that the limits of ultrasonic inversion are governed by the geometric structure of the forward map and the information content available in the measurement.

\begin{figure}
\centering
\includegraphics[width=0.4\linewidth]{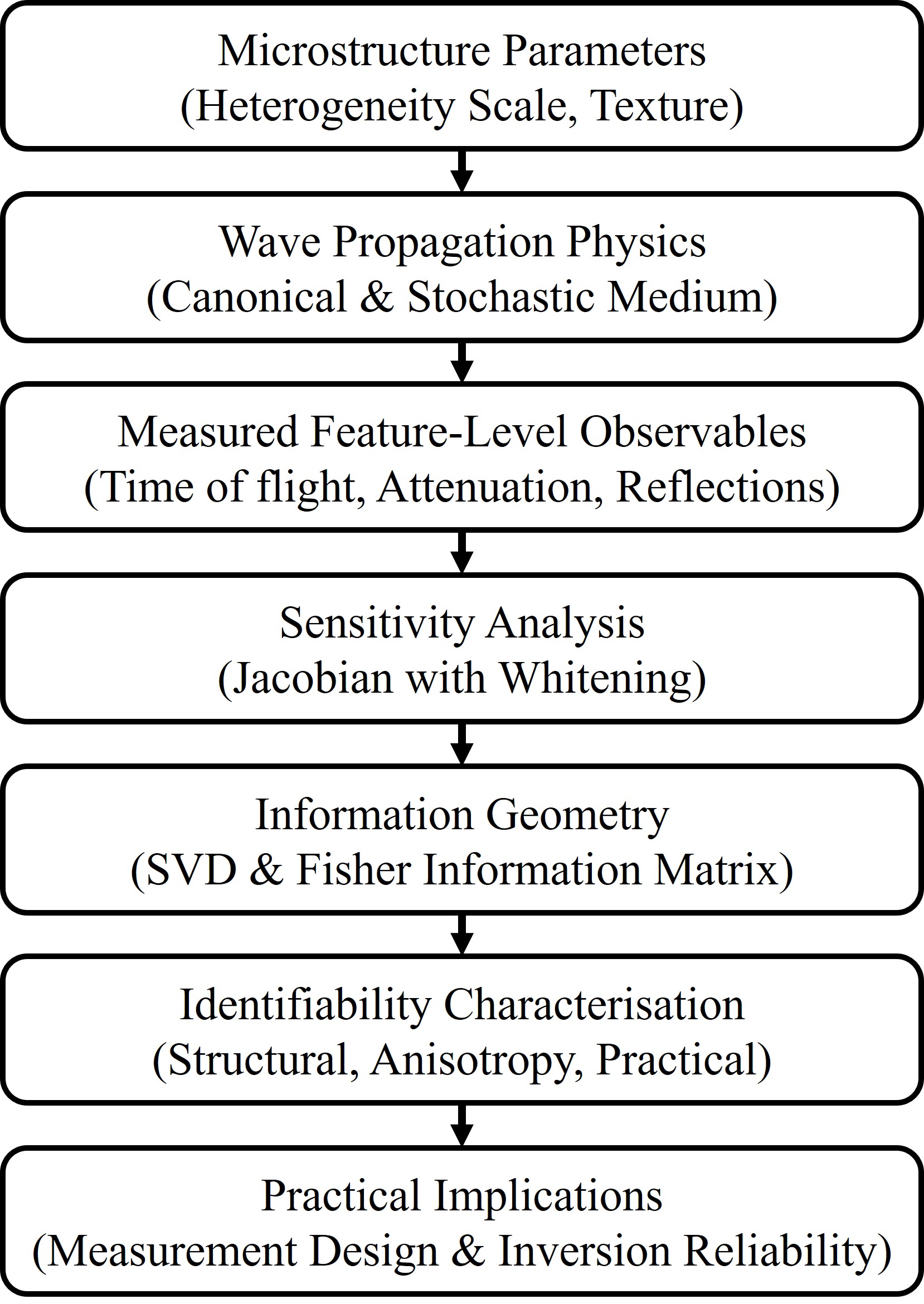}
\caption{Schematic of the identifiability analysis framework. Microstructure parameters are mapped to feature-level observables through wave propagation in the stochastic media. Sensitivity is quantified via the Jacobian, which is normalised using the covariance of stochastic microstructure variability. The resulting information geometry, characterised by singular values and the Fisher Information Matrix, determines structural and practical identifiability. These metrics in turn govern inversion reliability and inform ultrasonic measurement design.}
\label{fig:pipeline}
\end{figure}

The contributions of this paper are threefold. First, a geometric baseline for identifiability limits is developed using a canonical ultrasonic model. Second, heterogeneous surrogate media are found to exhibit full-rank but anisotropic sensitivity. Third, practical resolution is governed by the interplay between structural sensitivity and intrinsic stochastic variance. A schematic of the identifiability analysis framework is illustrated in Figure~\ref{fig:pipeline}.

\section{Canonical Model and Identifiability Framework}
\label{sec:framework}

\subsection{Model Configuration and Parameterisation}
\label{subsec:canonical_setup}

\begin{figure}
\centering
\includegraphics[width=0.6\linewidth]{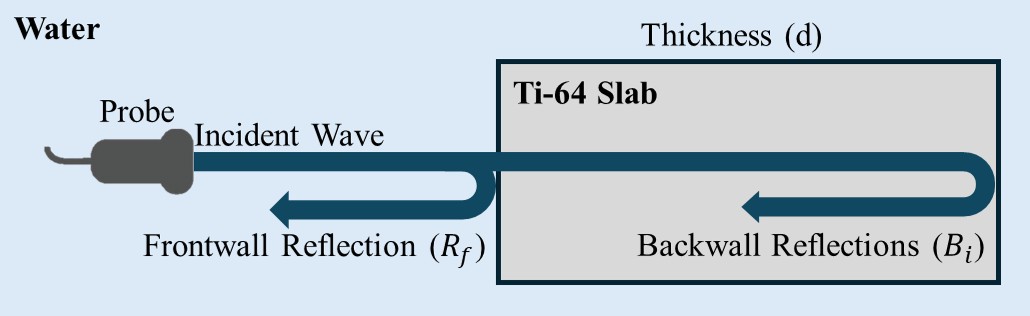}
\caption{An illustration of the canonical model configuration with a Ti-64 slab with unknown material properties embedded in water.}
\label{fig:Canonical}
\end{figure}

A canonical model is first used to represent a simple yet physically meaningful setup to assess identifiability limits before introducing microstructural complexity and stochastic variability \cite{Rose2014}.  The configuration is illustrated in Figure \ref{fig:Canonical}, in which a homogeneous solid titanium block of thickness $d$ is embedded in water. The properties of the canonical solid block are represented by the parameter vector
\begin{equation}
\mathbf{P} =
\begin{bmatrix}
\rho \\
d \\
\alpha \\
Z
\end{bmatrix},
\label{eq:P}
\end{equation}
where $\rho$ is the density, $d$ is the thickness, $\alpha$ is an effective attenuation coefficient which represents a lumped description of absorption and scattering losses, and $Z$ is the acoustic impedance, defined as the product of density $\rho$ and wave speed $v$.

In this one-dimensional canonical setting, wave propagation is treated in its simplest form, neglecting effects such as beam spread and mode conversion. Identifiability is assessed through a small set of physically interpretable, echo-indexed features that are consistent with experimental practice, where arrival times and echo amplitudes are more robust and transferable than full waveform inversion. The feature map is defined as
\begin{equation}
\mathbf{g}(\mathbf{P}) =
\begin{bmatrix}
\tau \\
\log\left|\dfrac{B_3}{B_2}\right| \\
\log|R_f|
\end{bmatrix},
\label{eq:g_P}
\end{equation}
where $\tau$ denotes the time-of-flight (TOF) between successive backwall reflections, providing a measure of wave speed. $B_2$ and $B_3$ are the complex amplitudes of the second and third backwall echoes, and the logarithmic ratio $\log|B_3/B_2|$ captures attenuation through cumulative energy loss. $R_f$ is the magnitude of the front-face reflection coefficient, encoding the impedance contrast between the fluid and solid. The corresponding expressions are
\begin{equation}
\tau = \frac{2d}{v} = \frac{2d\rho}{Z}, \qquad
\log\left|\frac{B_3}{B_2}\right| = \log|R_b| - 2\alpha d, \qquad
R_f = \frac{Z-Z_w}{Z+Z_w}, \quad
R_b = \frac{Z_w-Z}{Z_w+Z}.
\label{eq:canonical_relations}
\end{equation}
Wave velocity is defined as $v = Z/\rho$ to avoid redundancy while retaining acoustic impedance as the governing quantity. The inter-echo amplitude ratio is defined from successive echoes, with attenuation captured through the exponential decay over the path length of $2d$. The amplitude ratio is treated as a direct observable rather than an effective attenuation coefficient to avoid additional parameter dependencies in the feature definition. The reflection coefficients correspond to a symmetric water--solid--water configuration. Hence, these features provide a compact physical representation of the pulse--echo response, capturing propagation, attenuation, and interface behaviour.

\subsection{Identifiability \& Sensitivity Analysis}

\subsubsection{Analytical Jacobian, Rank Structure, and Parameter Geometry}

The sensitivity of the derived observable feature vector $\mathbf{g}(\mathbf{P})$ to changes in the canonical parameter vector $\mathbf{P}$ is represented by the Jacobian matrix,
\[
\mathbf{J} = \frac{\partial \mathbf{g}}{\partial \mathbf{P}},
\]
which measures the change in observable features with respect to a change in the parameter set. The closed-form expressions for the partial derivatives are summarised in Table~\ref{tab:jacobian} which highlights the physical structure of the forward map: TOF couples density and thickness, attenuation is predominant only in $\alpha d$, and interface observables depend solely on impedance contrast.

\begin{table}[htbp]
\centering
\caption{Analytical Jacobian for the canonical derived features.}
\label{tab:jacobian}
\begin{tabular}{lcccc}
\toprule
 & $\partial_\rho$ & $\partial_d$ & $\partial_\alpha$ & $\partial_Z$ \\
\midrule
$\tau = \dfrac{2d\rho}{Z}$
& $\dfrac{2d}{Z}$
& $\dfrac{2\rho}{Z}$
& $0$
& $-\dfrac{2d\rho}{Z^2}$ \\[6pt]
$\log\left|\dfrac{B_3}{B_2}\right|$
& $0$
& $-2\alpha$
& $-2d$
& $\dfrac{1}{R_b}\dfrac{\partial R_b}{\partial Z}$ \\[6pt]
$\log|R_f|$
& $0$
& $0$
& $0$
& $\dfrac{1}{R_f}\dfrac{\partial R_f}{\partial Z}$ \\
\bottomrule
\end{tabular}
\end{table}
Since the feature vector contains three observables while the parameter set has four unknowns, the Jacobian is dimensionally underdetermined. This implies that even under idealised noise-free conditions, no more than three independent parameter combinations can be identified locally. To account for this rank deficiency, an additional reduced configuration is considered
\[
\mathbf{P}_C = [\alpha, Z]^T
\]
with density and thickness of the canonical slab treated as known parameters. In this scenario, the Jacobian becomes full column rank and this helps to separate two distinct phenomena: (i) dimensional non-identifiability arising from rank deficiency, and (ii) conditioning degradation arising from structural parameter coupling within the identifiable subspace.

\subsubsection{Reference Configuration: Singular Values and Conditioning}

Identifiability is generally characterised through the singular value decomposition of $\mathbf{J}$. The singular values $\{\sigma_i\}$ quantify the local deformation of parameter space under the mapping $\mathbf{g}(\mathbf{P})$ and therefore indicate the strength of parameter combinations that are observable through the measured features. The number of non-zero singular values is equal to the rank of $\mathbf{J}$. Large singular values correspond to dominant modes that induce strong and distinguishable changes in the ultrasonic response, while small singular values indicate weakly observable combinations that are difficult to resolve. The following condition number
\[
\mathrm{cond}(\mathbf{J}) = \frac{\sigma_{\max}}{\sigma_{\min}}
\]
measures the anisotropy of the local information geometry. A large condition number represents an elongated information ellipsoid and strong parameter coupling, while an ideal value of 1 indicates a balanced and well-conditioned inverse problem.

The Jacobian for the canonical setup is evaluated at a reference Ti-64--water configuration using the analytical expressions in Table~\ref{tab:jacobian}. Ti-64 material properties are: $\rho = 4500~\mathrm{kg\,m^{-3}}$, $c = 62400~\mathrm{m\,s^{-1}}$ (giving $Z = 28.08~\mathrm{MRayl}$), attenuation $\alpha = 20~\mathrm{Np\,m^{-1}}$, and thickness $d = 10~\mathrm{mm}$. The impedance for water is $Z_w = 1.48~\mathrm{MRayl}$. Singular values are computed from the log-parameter scaled Jacobian and presented in Table~\ref{tab:scaled_svs_reference}.

\begin{table}[htbp]
\centering
\caption{Singular values of the log-parameter scaled Jacobian at the Ti-64--water reference configuration. Case~1 corresponds to the full parameterisation $\mathbf{P}=[\rho,d,\alpha,Z]^T$, and Case~2 corresponds to the reduced parameterisation $\mathbf{P}_C=[\alpha,Z]^T$.}
\label{tab:scaled_svs_reference}
\begin{tabular}{ccccc}
\toprule
Case & $\sigma_1$ & $\sigma_2$ & $\sigma_3$ & $\mathrm{cond}(J)$ \\
\midrule
1
& $1.147 \times 10^{0}$ 
& $1.818 \times 10^{-1}$ 
& $8.165 \times 10^{-6}$ 
& $1.404 \times 10^{5}$ \\
2
& $8.22 \times 10^{-1}$ 
& $1.793 \times 10^{-1}$ 
& -- 
& $4.584 \times 10^{0}$ \\
\bottomrule
\end{tabular}
\end{table}

For Case 1, three non-zero singular values are generated which is consistent with the Jacobian rank of 3. The dominant singular values ($\sigma_1$ and $\sigma_2$) contain most of the recoverable information and are associated with perturbation modes that strongly influence reflection amplitude and attenuation. The smallest singular value ($\sigma_3$) corresponds to a near-degenerate parameter direction due to structural coupling between density and thickness. This is also reflected in the large conditioning number of $\approx 1.4\times 10^{5}$. Although the forward map is not full rank over the full four-parameter space, the observables still contain information about three independent parameter combinations. Though due to the presence of a null-direction, the inversion is non-unique and unstable.

In contrast, the reduced system in Case 2 yields two singular values of comparable magnitude. While both values are slightly lower than those in Case~1, the overall condition number for Case 2 is closer to unity. This indicates a nearly isotropic information ellipsoid and a stable inversion problem, mainly due to the removal of the coupled parameters density and thickness. Hence, the singular value spectrum shows that identifiability depends not only on dimensionality, but also on how evenly sensitivity is distributed across parameter directions.

\subsubsection{Sensitivity Variation with Impedance Contrast}

\begin{figure}
\centering
\includegraphics[width=0.6\linewidth]{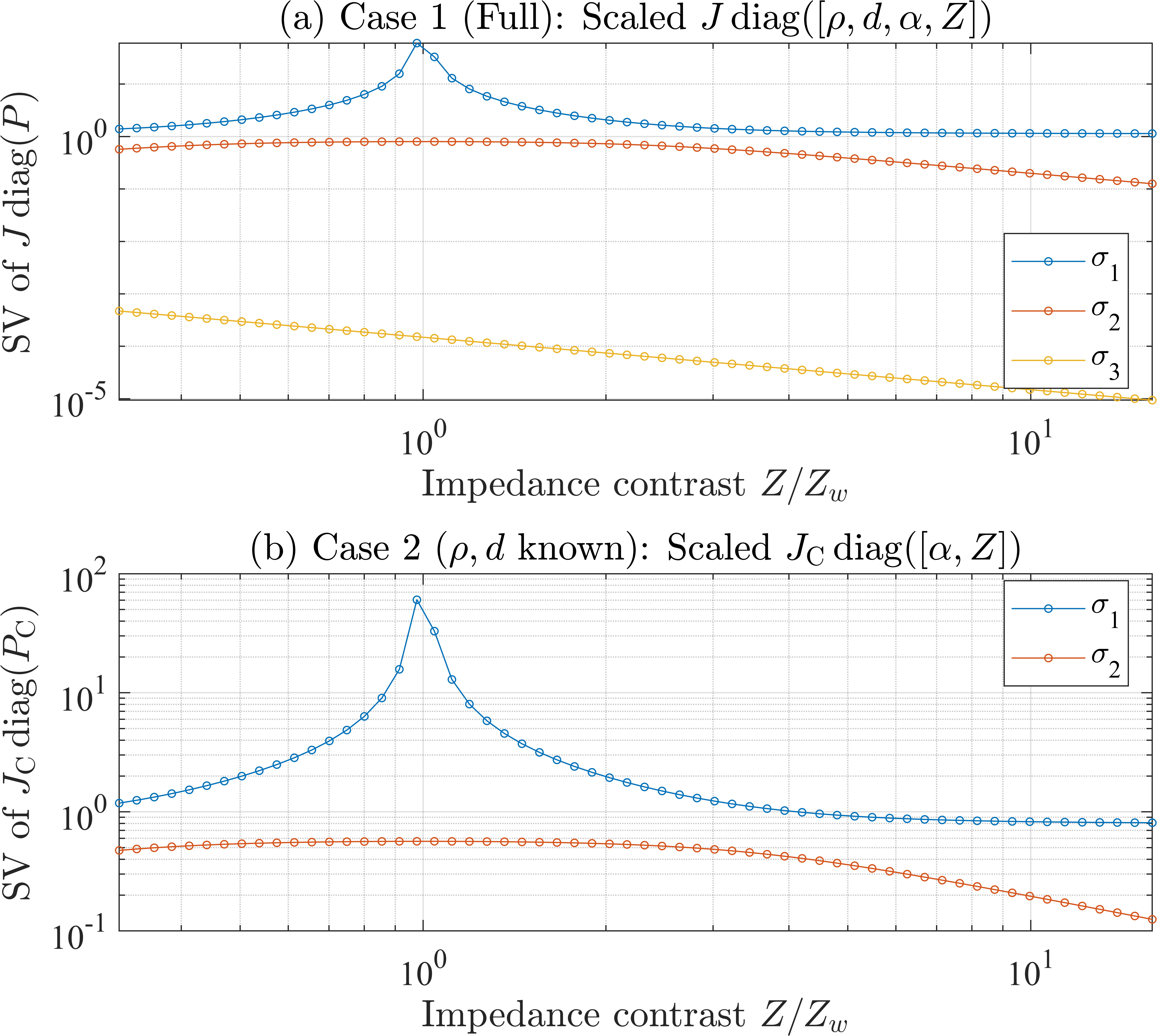}
\caption{
Singular value spectra of the log-parameter scaled Jacobian as a function of impedance contrast $Z/Z_w$. 
(a) Full parameterisation ($\mathbf{P}=[\rho,d,\alpha,Z]^T$) exhibiting persistent spectral separation and a near-null direction associated with structural parameter coupling. 
(b) Reduced parameterisation ($\mathbf{P}_C=[\alpha,Z]^T$) in which the weakly observable mode is removed, yielding balanced sensitivity within the identifiable subspace. 
The contrast-dependent peak near $Z/Z_w\approx1$ reflects enhanced interface sensitivity, while decay at high contrast indicates reflection saturation.
}
\label{fig:sv_vs_contrast}
\end{figure}

To examine how identifiability changes with material variations, impedance contrast $Z/Z_w$ is varied as it influences all three observables. This isolates the interface-driven sensitivity and enables a direct examination of saturation effects that arise at high contrast when $|R_f|$ tends to one. The results for Cases 1 and 2 are compared in Figure~\ref{fig:sv_vs_contrast}, illustrating how the singular values vary with respect to impedance contrast.

In both the (a) full and (b) reduced systems, a peak near $Z/Z_w \approx 1$ is observed for dominant singular value $\sigma_1$. This reflects enhanced local sensitivity in the impedance-matching regime, where small changes in $Z$ produce large relative changes in $R_f$ and $R_b$. Physically, reflection amplitude diminishes as impedance mismatch decreases and tends to zero near impedance-matching. However, the reflection coefficient varies most rapidly in this regime, so the corresponding gradients are maximised, leading to increased Jacobian sensitivity despite weak signals. This indicates that sensitivity does not necessarily translate to practical observability.

In the full system (a), the persistence of strong singular-value separation across the sweep shows that ill-conditioning is not confined to a specific operating point, but is instead a structural feature of the parameterisation. In addition, the smallest singular value decreases progressively with increasing impedance contrast, indicating a loss of independent parameter sensitivity. In contrast, the reduced system (b) exhibits two singular values of comparable magnitude throughout, showing that removal of the structurally coupled $(\rho,d)$ direction eliminates the near-null mode responsible for the extreme anisotropy in Case~1 and restores a more balanced and well-conditioned inverse problem.

\begin{figure}
\centering \includegraphics[width=0.55\linewidth]{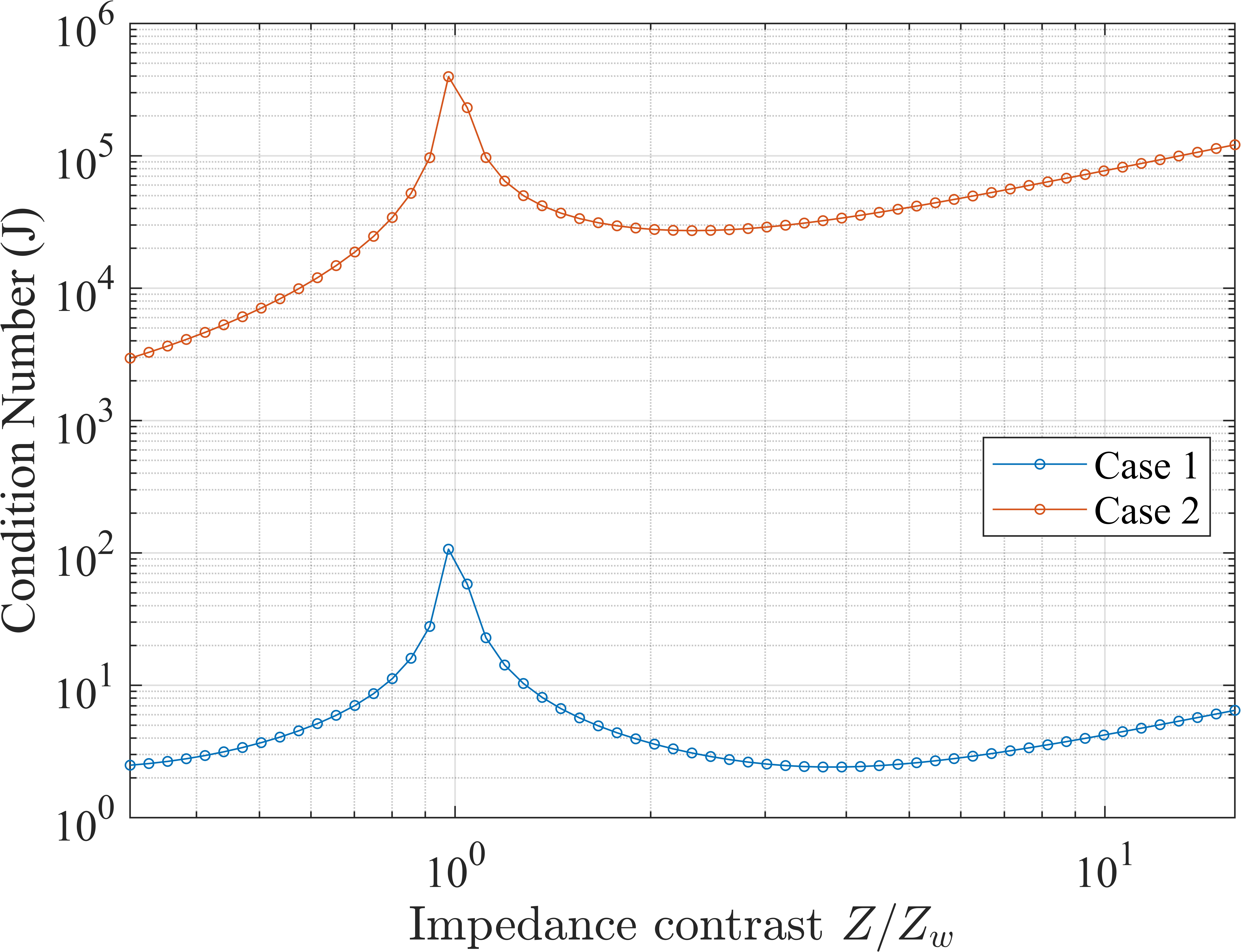} 
\caption{Jacobian condition number versus impedance contrast $Z/Z_w$ for the full (Case 1) and reduced (Case 2) parameterisations. The strong separation between the two curves shows that density--thickness coupling in Case 1 produces a near-null direction and severe ill-conditioning.}
\label{fig:cond_vs_contrast}
\end{figure}

The corresponding condition numbers for both cases with respect to impedance contrast are compared in Figure \ref{fig:cond_vs_contrast}. Even though the two profiles are visually similar, they exhibit fundamentally different geometric behaviour. For Case 1, the condition number remains large across the entire impedance contrast range, with a pronounced peak at $Z/Z_w \approx 1$ that corresponds to the impedance-matching regime. In this regime, small perturbations in impedance produce large relative changes in the reflection coefficients and temporarily enhance sensitivity. This stretches the information geometry along impedance-dominated directions, but the associated improvement is limited by the unresolved density–thickness coupling.

As impedance contrast increases further, the condition number rises rapidly. This is driven by saturation of the reflection coefficients as $|R_f|$ tends to one, where further changes in impedance produce only weak variations in $\log|R_f|$. However, Case~2 remains relatively well-posed as the condition number remains within two orders of magnitude throughout the impedance sweep, as a result of the resolved density–thickness coupling and the elimination of the near-null mode.

\subsection{Finite-Difference Validation of Canonical Features}
\label{subsec:fd_validation}

The canonical results are validated using a one-dimensional finite-difference time-domain (FDTD) model with the scalar acoustic wave equation \cite{Alford1974,Zakaria2003,Schubert2004}. To emulate the canonical setup, a homogeneous solid titanium block between two water regions is simulated, and a normal-incident broadband Ricker wavelet is generated from the water surface. The signals are recorded at a fixed location on the water-solid interface, with TOF estimated based on the arrival times between the second and third back wall reflections, and the front-face reflection magnitude taken as peak amplitude for comparison.

\begin{figure}
\centering
\includegraphics[width=1\linewidth]{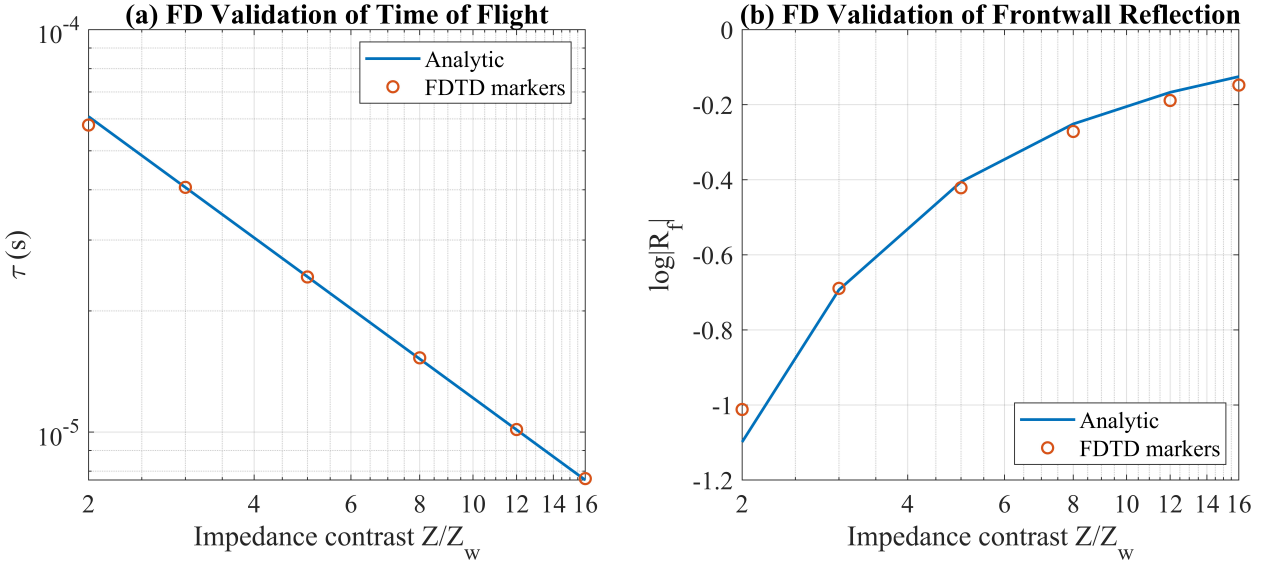}
\caption{
Finite-difference validation of canonical results comparing (a) Time-of-flight $\tau$ versus impedance contrast and (b) log-magnitude of the front-wall reflection coefficient $\log|R_f|$ versus impedance contrast. The solid curves represents the canonical results, and markers denote the FDTD measurements, with good agreement between them.}
\label{fig:fd_validation}
\end{figure}

The results are illustrated in Figure~\ref{fig:fd_validation} which compares both the FDTD and canonical responses across a range of impedance contrasts. In general, good agreement is observed for both (a) TOF and (b) reflection magnitude, with small deviations at extreme contrast due to poorer parameter sensitivity. The linear dependence of TOF and the non-linear variation of frontwall reflection with impedance contrast are also reproduced by the numerical model, indicating good agreement and confirming that the canonical feature relations capture the appropriate dependencies between the measured observables and material parameters.

\subsection{Canonical Insights and Implications for Microstructural Models}

The canonical analysis reveals three distinct features of the inverse problem. First, when the number of unknowns exceeds the number of independent observables, recoverability is limited at the dimensional level as not all parameter directions can be resolved. Second, even within the identifiable subspace, parameter coupling introduces anisotropy which produces strong singular-value separation and poor conditioning. Third, forward-map saturation reduces sensitivity when the observables become weakly responsive to parameter changes, thereby diminishing the amount of recoverable information. Hence, these results show that non-uniqueness and instability exist in a simple homogeneous setting. The canonical model therefore serves as a baseline for separating dimensional restriction, parameter coupling, and interface saturation before examining how related effects appear in heterogeneous stochastic media \cite{AnstettCollin2020}.

\section{Surrogate Microstructure Models}
\label{sec:microstructure}

Having established the intrinsic identifiability structure of a canonical homogeneous model, the feature-level framework is extended to heterogeneous media represented by statistically generated microstructure surrogates. The objective is to examine the wave information geometry in heterogeneous media.

\subsection{Methodology}

\subsubsection{Microstructure Parameterisation and Model Setup}
\label{subsec:microstructure_params}

\begin{figure}
\centering
\includegraphics[width=1\linewidth]{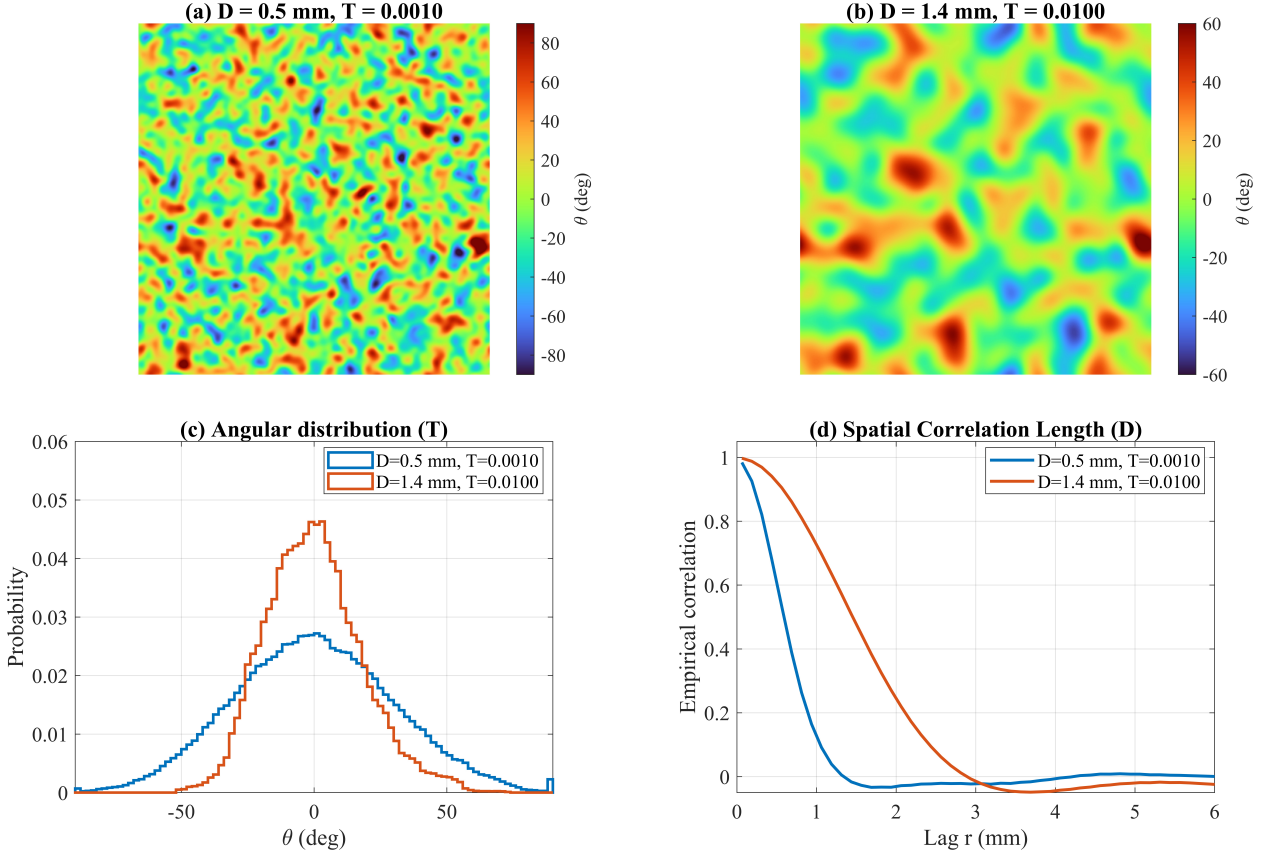}
\caption{Examples of surrogate microstructures generated using Gaussian random fields.
(a) Low correlation length and low texture-coherence parameter ($D=0.5$ mm, $T=0.001$), showing fine-scale heterogeneity and broadly distributed orientations.
(b) High correlation length and higher texture-coherence parameter ($D=1.4$ mm, $T=0.01$), showing coarse heterogeneity with stronger orientation coherence.
(c) Orientation angular distributions based on different texture-coherence $T$ values. (d) Empirical spatial correlation functions for different correlation lengths $D$.}
\label{fig:GRF1}
\end{figure}

Microstructural heterogeneity is represented using statistically generated surrogate models based on Gaussian random fields (GRFs) \cite{Stanke1984a,Kube2015,Liu2019a}. Rather than explicitly resolving individual grains, the medium is modelled as a spatially varying elastic stiffness field obtained by assigning locally varying crystallographic orientations to a baseline Ti--6Al--4V stiffness tensor taken as $C_{11}=162$ GPa, $C_{33}=181$ GPa, $C_{12}=92$ GPa, $C_{13}=69$ GPa, $C_{44}=46.7$ GPa, and $C_{66}=35$ GPa \cite{Lan2015c}. The surrogate microstructure is parameterised by two descriptors: a correlation length $D$, which sets the spatial scale of heterogeneity, and a texture-coherence parameter $T$, which controls the smoothness and angular spread of the orientation field. Let $\xi(x,z)$ denote a zero-mean, unit-variance GRF defined on the computational mesh with isotropic covariance
\[
\mathbb{E}\big[\xi(\mathbf{x})\,\xi(\mathbf{x}')\big]
=
C\!\left(\frac{\|\mathbf{x}-\mathbf{x}'\|}{D}\right),
\]
The GRF is mapped to a local orientation field $\theta(x,z)$ through a transformation controlled by $T$, such that increasing $T$ produces a smoother and more spatially coherent orientation distribution with reduced local angular fluctuations. In the present two-dimensional setting, $\theta(x,z)$ represents an in-plane rotation angle of the local crystallographic $c$-axis, so that the material anisotropy varies only through rotations within the $(x,z)$ plane. The elastic stiffness tensor at each point is then obtained by rotating a reference Ti--6Al--4V stiffness tensor $\mathbf{C}_0$ according to the local orientation,
\[
\mathbf{C}(x,z) = \mathbf{R}(\theta(x,z))\, \mathbf{C}_0\, \mathbf{R}^\top(\theta(x,z)),
\]
where $\mathbf{R}(\theta)$ denotes the rotation operator. This construction enables controlled variation of both the spatial scale and texture coherence of microstructural heterogeneity, while remaining computationally tractable for systematic parameter studies. This parameterisation is illustrated in Figure \ref{fig:GRF1} with two different cases, (a) representing a low-D and low-T system which contains small spatial variations and more evenly distributed crystal orientations that range up to 90$^{\circ}$, and (b) represents a high-D and high-T system which contains large spatial variations and preferred crystal orientations that range up to 60$^{\circ}$.

The two-dimensional domain of 30 by 30 mm with a mesh size of 0.05 mm is imported directly for plane-wave simulations performed with the finite-element (FE) solver Pogo \cite{Huthwaite2014}. An incident plane longitudinal wave is excited from the top surface of the solid medium, supported by symmetrical boundary conditions on both sides. The FE simulations are conducted at a centre frequency of 10~MHz over a pre-defined range and combinations of correlation length $D$ and texture-coherence parameter $T$ listed in Table \ref{tab:grf_params}, with a total of 100 combinations, each with 40 statistically independent realisations by randomising the seed points. This number was chosen as a practical compromise between statistical stability and computational cost, and is consistent with prior studies showing convergence of averaged attenuation metrics with a similar number of realisations \cite{Huang2020}. With other material properties held constant, wave perturbations therefore arise solely from spatial fluctuations in elastic stiffness due to variations of $D$ and $T$.

\begin{table}
\centering
\caption{Parameter ranges used for the GRF surrogate microstructure simulations.}
\begin{tabular}{c c}
\toprule
Parameter & Values \\
\midrule
Correlation length $D$ (mm) 
& $[0.5,\;0.6,\;0.7,\;0.8,\;0.9,\;1.0,\;1.1,\;1.2,\;1.3,\;1.4]$ \\

Texture-coherence parameter $T \times 10^{-3}$ 
& $[1.0,\;2.0,\;3.0,\;4.0,\;5.0,\;6.0,\;7.0,\;8.0,\;9.0,\;10]$ \\
\bottomrule
\end{tabular}
\label{tab:grf_params}
\end{table}

By construction, $(D,T)$ are surrogate descriptors rather than literal metallographic grain size or crystallographic texture parameters. They provide a minimal two-parameter representation of heterogeneity scale and orientation-field coherence that suffices to examine how ultrasonic velocity and attenuation observables respond to controlled microstructural variability \cite{Liu2019a}. The identifiability limits derived here are therefore conditional on the chosen feature representation and should be interpreted as limits associated with practically accessible observables, rather than as universal limits of the full waveform response.

\subsubsection{Simulation Results}

With the surrogate microstructure defined, the analysis focuses on the two observables that can be extracted from the simulated responses---wave attenuation and velocity. For attenuation, the amplitude ratio between the input and back reflection signals is used. Since the model comprises a single medium with plane-wave propagation, the transmission/reflection coefficients and diffraction correction factor are neglected. This is represented by:
\begin{equation} \label{att}
\alpha_s = \frac{1}{z} \ln\left( \frac{F}{B} \right)
\end{equation}
where $z$ is the propagation distance, and $F$ and $B$ are the front and back wall signal amplitudes in the frequency domain, respectively \cite{VanPamel2015a}. For ultrasound velocity, the cross-correlation method is used---the similarity between two signals is measured as the latter is shifted in time \cite{Zhang2008}. The time shift at which the best match occurs is the TOF, which, divided by the propagation distance, gives the wave velocity. These two observables define the surrogate feature vector,
\[
\mathbf{g}(D,T)=
\begin{bmatrix}
v(D,T)\\
\alpha(D,T)
\end{bmatrix},
\]
The corresponding mean feature map and intrinsic microstructure covariance over the ensemble of stochastic realisations are
\[
\boldsymbol{\mu}(D,T)=\mathbb{E}_r\!\left[\mathbf{g}^{(r)}(D,T)\right],
\qquad
\boldsymbol{\Sigma}_{\mathrm{ms}}(D,T)=\mathrm{Cov}_r\!\left(\mathbf{g}^{(r)}(D,T)\right).
\]
where $\boldsymbol{\mu}(D,T)$ represents the deterministic component of the forward operator, while $\boldsymbol{\Sigma}_{\mathrm{ms}}(D,T)$ quantifies the variability arising from stochastic microstructure realisations.

\begin{figure}
\centering
\includegraphics[width=0.9\linewidth]{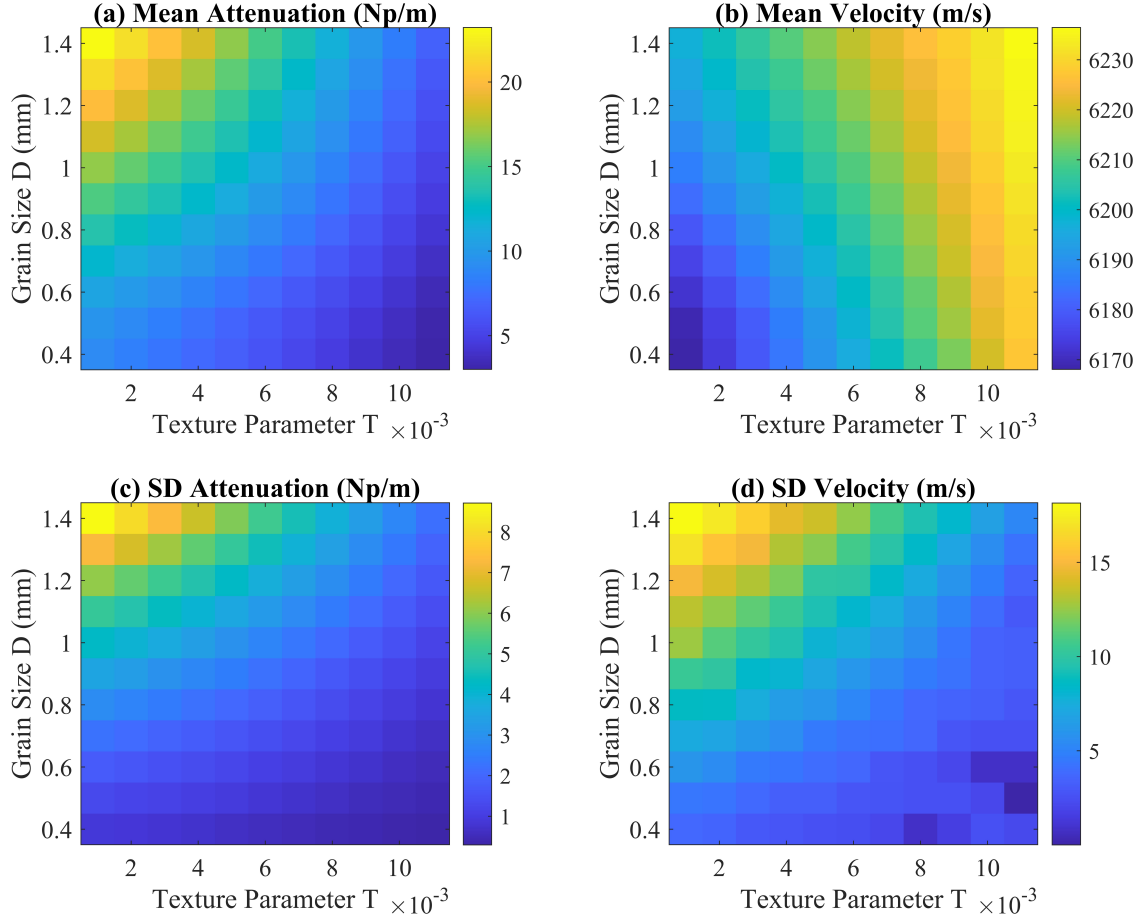}
\caption{Mean and variability of ultrasonic observables obtained from Gaussian random field (GRF) surrogate microstructures across the correlation length ($D$) and texture-coherence parameter ($T$) space across 40 realisations. (a) Mean attenuation $\alpha$, (b) mean longitudinal wave velocity $v$, (c) standard deviation of attenuation, and (d) standard deviation of velocity computed at each $(D,T)$ combination.}
\label{fig:GRF2}
\end{figure}

The corresponding simulated attenuation and velocity results across the correlation length ($D$) and texture-coherence ($T$) parameter space are illustrated in Figure~\ref{fig:GRF2}. The mean attenuation map in (a) shows a strong increase in attenuation with increasing correlation length, consistent with the expected rise in scattering as the characteristic microstructural length scale approaches the ultrasonic wavelength. At the same time, attenuation decreases systematically with increasing $T$. In the GRF representation used here, larger values of $T$ correspond to increased spatial correlation and smoother orientation fields, which reduce local elastic contrast and therefore weaken scattering strength. As a result, attenuation is primarily governed by correlation length while being modulated by the texture-coherence parameter \cite{Bai2020}.

The mean velocity map in Figure~\ref{fig:GRF2}(b) exhibits a different dependence on the GRF parameters. Velocity varies predominantly with the texture-coherence parameter, increasing monotonically with $T$, while showing a weaker dependence on correlation length. This behaviour reflects the fact that wave velocity is primarily determined by the effective elastic stiffness of the medium, which is strongly influenced by the orientation distribution of grains \cite{Lan2015a} but only weakly affected by the spatial scale of the microstructure \cite{Ming2020}. Consequently, the velocity observable provides stronger sensitivity to texture as opposed to correlation length.

The standard deviation (SD) maps in Figure~\ref{fig:GRF2}(c)–(d) quantify the variability of the observables across the ensemble of GRF realisations. The SD of attenuation increases significantly with correlation length $D$ and decreases with increasing texture-coherence parameter $T$. This trend reflects the increasing importance of scattering fluctuations in regimes where the microstructural scale becomes comparable to the wavelength, resulting in stronger variability between individual realisations of the random microstructure. A similar but weaker pattern is observed for the velocity SD in (d), which also increases in regimes associated with stronger scattering. In both observables, the variability therefore follows the overall scattering strength of the medium.

Hence, these results highlight the complementary sensitivities of the two observables: attenuation is more strongly influenced by $D$ while remaining modulated by $T$, while velocity has stronger sensitivity to $T$. At the same time, the variability maps demonstrate that the measurement uncertainty associated with each observable is not uniform across the parameter space but depends on the scattering intensity. These trends form the basis for the subsequent identifiability analysis, where the sensitivity and variability of the observables are combined to evaluate the recoverability of the underlying microstructure parameters.

While the observable maps in Figure~\ref{fig:GRF2} provide qualitative insights into how attenuation and velocity vary across the $(D,T)$ parameter space, these trends alone do not determine whether the microstructure parameters can be uniquely recovered from the measurements. In practice, successful inversion depends not only on how strongly the observables change with the parameters, but also on how independently the parameters influence the measured responses and how sensitive those responses are relative to the variability introduced by the stochastic microstructure.

\subsection{Identifiability and Sensitivity Analysis}
\label{subsec:microstructure_identifiability}

Local sensitivities of the ensemble-mean features with respect to $(D,T)$ are evaluated using central finite differences \cite{Li2017}. For each nominal grid point $(D_i,T_j)$, the simulation results at the nearest intervals in $(D_i \pm \Delta D, T_j)$ and $(D_i, T_j \pm \Delta T)$ are used to compute the gradient. Common seed points are used across the parameter sweeps to reduce Monte Carlo noise. The Jacobian of the mean feature map,
\[
\mathbf{J}_\mu(D,T)
=
\frac{\partial \boldsymbol{\mu}}{\partial (D,T)},
\]
is approximated component-wise as
\[
\frac{\partial \mu_k}{\partial D}
\approx
\frac{\mu_k(D+\Delta D,T)-\mu_k(D-\Delta D,T)}{2\Delta D},
\qquad
\frac{\partial \mu_k}{\partial T}
\approx
\frac{\mu_k(D,T+\Delta T)-\mu_k(D,T-\Delta T)}{2\Delta T},
\]
where $\mu_k$ denotes the $k$-th component of $\boldsymbol{\mu}$. As in the canonical case, this matrix characterises how perturbations in the parameters map into changes in the observables. This procedure yields a $2\times2$ numerical Jacobian at each parameter grid point. Since the number of observables $(\alpha,v)$ matches the number of surrogate descriptors $(D,T)$, the forward map is locally square. Hence, identifiability in this surrogate configuration is not limited by observable count, and any loss must arise from anisotropic conditioning of the Jacobian or from the stochastic variability in the microstructure.

\subsubsection{Structural Identifiability}

To describe the sensitivity structure of the surrogate model, several complementary metrics are evaluated from the ensemble-mean Jacobian, as summarised in Table~\ref{tab:identifiability_metrics}. The singular values (SV) indicate the strength of the independent parameter directions, where larger values mean that changes in the parameters produce stronger changes in the observables. The condition number measures the sensitivity anisotropy across the parameter directions. The determinant indicates the total information content available in the system. For interpretation, the Jacobian is examined in raw, normalised, and whitened forms. The raw Jacobian reflects absolute physical sensitivity, the normalised Jacobian removes observable scale effects, and the whitened Jacobian measures sensitivity relative to variability.

In particular, the raw and normalised Jacobians are first examined to understand the identifiability structure. The raw Jacobian reflects the absolute sensitivity of the observables to the surrogate parameters and is influenced by the underlying physics as well as the magnitude and units of the measured quantities. Hence, an observable may appear to dominate simply because its raw variation is larger. The normalised Jacobian removes this scaling effect and instead reflects proportional sensitivity. In this form, it provides a scale-independent view of the local sensitivity structure by showing how percentage changes in the parameters are transferred into percentage changes in the observables.

\begin{table}
\centering
\caption{Sensitivity-based metrics used to assess structural identifiability of the surrogate model.}
\begin{tabular}{lll}
\toprule
\textbf{Step} & \textbf{Metric} & \textbf{Interpretation} \\
\midrule
Jacobian & $\mathbf{J}_\mu = \partial \boldsymbol{\mu} / \partial (D,T)$ 
& Local sensitivity of observables to parameters \\

Singular values & $S_i = \sigma_i(\mathbf{J}_\mu),\; i=1,2$ 
& Strength of independent parameter directions \\

Condition number & $\mathrm{cond}(\mathbf{J}_\mu)=S_1/S_2$ 
& Sensitivity anisotropy (directional imbalance) \\

Determinant & $\det(\mathbf{J}_\mu)=S_1*S_2$ 
& Joint sensitivity magnitude (information content) \\

\bottomrule
\end{tabular}
\label{tab:identifiability_metrics}
\end{table}

\begin{figure}
\centering
\includegraphics[width=0.8\linewidth]{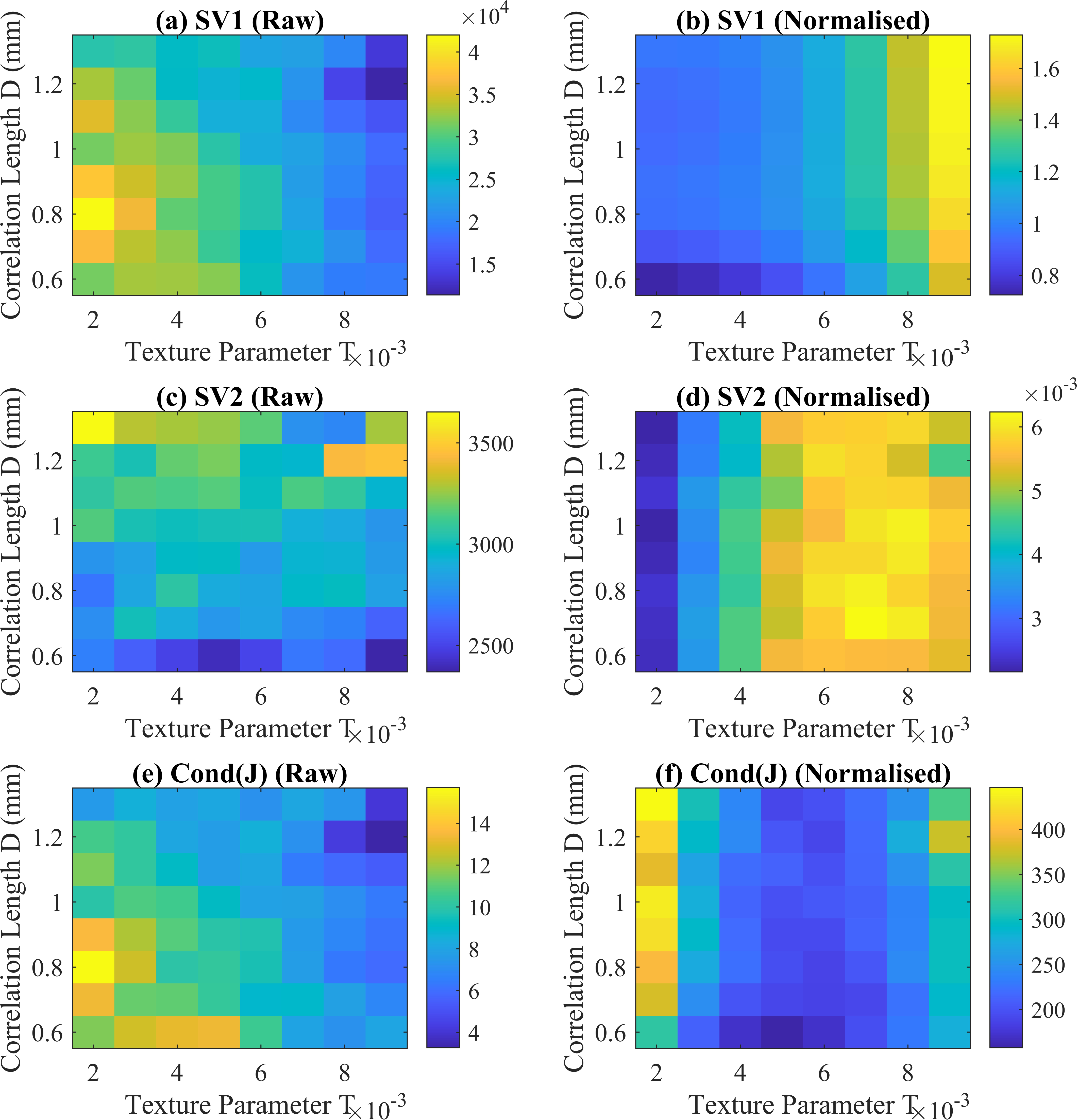}
\caption{
Comparison of raw and normalised sensitivity metrics across the $(D,T)$ domain. (a)–(b) shows the largest singular value (SV1), (c)–(d) the smallest singular value (SV2), and (e)–(f) the corresponding condition number of the Jacobian. The raw Jacobian reflects absolute sensitivity of the observables, while the normalised Jacobian represents proportional (scale-independent) sensitivity. The maps illustrate how observable scaling influences both the strength and anisotropy of the local sensitivity structure.}
\label{fig:GRF3}
\end{figure}

The raw and normalised Jacobian metrics are compared in Figure~\ref{fig:GRF3} across the $(D,T)$ domain. The largest singular value in Figure~\ref{fig:GRF3}(a)-(b) corresponds to the dominant local sensitivity direction. In the raw Jacobian, SV1 is strongest at lower correlation length $D$ and lower texture-coherence parameter $T$, where the observables undergo the largest absolute changes. This behaviour is consistent with the stronger absolute variation of the attenuation observable in that region, reflecting stronger scattering-driven contrast in the response. After normalisation, the stronger region shifts towards higher $T$ values, showing that the dominant proportional sensitivity is more closely related to variation in $T$. In this form, both attenuation and velocity contribute more comparably, and the dominant direction reflects the stronger relative dependence of the observables on $T$.

The smallest SV presented in Figure~\ref{fig:GRF3}(c)-(d) corresponds to the weaker independent parameter direction. In the raw Jacobian, SV2 remains finite across the interior of the domain, indicating that the forward map stays full rank and that the two surrogate parameters remain structurally distinguishable. This weaker mode is consistent with the more limited sensitivity of the observables to correlation length $D$, since most of the $D$-dependence enters through attenuation while velocity responds more strongly to texture-coherence parameter $T$. Consequently, the secondary direction is preserved, but with a smaller absolute strength than the dominant $T$-related direction. In the normalised case, SV2 becomes larger over a broad central to high-$T$ region, showing that once the observable scale is removed, the weaker $D$-sensitive direction is retained more clearly in proportional terms. This indicates that part of the anisotropy seen in the raw Jacobian arises from unequal observable scaling, rather than from directional collapse of the forward map.

Lastly, the condition numbers for the raw and normalised Jacobians are compared in Figure~\ref{fig:GRF3}(e)-(f). In the raw Jacobian, the condition number remains moderate over most of the domain. Although the two singular directions are unequal, the absolute sensitivity structure is still well-conditioned, and the chosen observables respond to both the correlation length $D$ and texture-coherence $T$ variation with sufficient absolute contrast. After normalisation, the condition number increases noticeably, especially at low $T$ and near the edges of the domain. This suggests that once the observable scale is removed, the relative sensitivity becomes more uneven, with the $T$-dominated direction contributing more strongly than the $D$-sensitive direction. In physical terms, the observables carry proportionally richer information about $T$ than about $D$ in these regions, because the influence of $D$ enters mainly through the weaker attenuation-sensitive channel. Hence, the rise in condition number reflects the increasing imbalance in how the two microstructural descriptors are encoded in the measurements.

Overall, the raw and normalised Jacobians show that the surrogate forward map remains full rank, but has an uneven sensitivity structure across the $(D,T)$ domain. The SV structure indicates that the dominant direction is associated mainly with variations in texture-coherence parameter $T$, while the weaker direction is predominantly linked to correlation length $D$. This is consistent with the structure of the forward map: velocity $v$ varies predominantly with $T$, whereas attenuation $\alpha$ responds to both $D$ and $T$ over the explored domain. The structural identifiability landscape is therefore shaped less by directional overlap than by an uneven weighting of the two parameter directions in the observable space.

This distinction is important because structural identifiability alone does not guarantee reliable recovery in practice. Although the two parameter directions remain locally distinguishable, the weaker direction is expressed less strongly in the observables than the dominant one, so the inverse problem is already anisotropic. If the sensitivity along this weaker direction becomes comparable to measurement uncertainty or microstructural variability, it may become effectively unrecoverable. This motivates the next step of the analysis, where the Jacobian is interpreted relative to the variability of the observables through a practical identifiability framework.

\subsubsection{Practical Identifiability}

Identifiability is additionally influenced by stochastic variability in the microstructure. As shown in Figure~\ref{fig:GRF2}(c)-(d), the observable responses exhibit intrinsic variance between realisations, so parameter recoverability depends not only on the magnitude of sensitivity, but on sensitivity relative to this variability. To account for this effect, the sensitivity analysis is extended to a variance-weighted whitened Jacobian, where local sensitivities are scaled by the inverse covariance of the observables \cite{Lam2022}. The corresponding metrics are summarised in Table~\ref{tab:practical_identifiability}. Structural metrics describe the geometry of the forward map, while variance-weighted metrics quantify the effective information content under microstructure variability.

\begin{table}[h]
\centering
\caption{Variance-weighted (practical) identifiability metrics incorporating intrinsic microstructure variability.}
\begin{tabular}{lll}
\toprule
\textbf{Step} & \textbf{Metric} & \textbf{Interpretation} \\
\midrule
Covariance & $\boldsymbol{\Sigma}_{\text{ms}}$ 
& Intrinsic variability of observables across microstructure realisations \\

Whitened Jacobian & $\tilde{\mathbf{J}} = \boldsymbol{\Sigma}_{\text{ms}}^{-1/2}\mathbf{J}_\mu$ 
& Sensitivity normalised by variability \\

Fisher information & $\mathbf{F} = \tilde{\mathbf{J}}^\top \tilde{\mathbf{J}}$ 
& Information content accounting for variability \\

Fisher eigenvalues & $\lambda_i(\mathbf{F})$ 
& Strength of resolvable parameter directions \\

Uncertainty bound & $\mathbf{F}^{-1}$ 
& Lower bound on parameter estimation uncertainty (CRLB) \\

Gradient orientation & $\theta(\tilde{\partial}_D \boldsymbol{\mu},\, \tilde{\partial}_T \boldsymbol{\mu})$
& Directional independence after variance weighting \\
\bottomrule
\end{tabular}
\label{tab:practical_identifiability}
\end{table}

An additional metric is the gradient orientation, which evaluates the directional alignment of the parameter sensitivities in feature space. In the variance-weighted setting, these directions are represented by the two columns of the whitened Jacobian, corresponding to perturbations in $D$ and $T$ projected into the observable space $(v,\alpha)$ after accounting for intrinsic variability. The alignment angle between these directions is defined as
\[
\theta
=
\cos^{-1}
\left(
\frac{
\langle \tilde{\mathbf{J}}_{:,1}, \tilde{\mathbf{J}}_{:,2} \rangle
}{
\|\tilde{\mathbf{J}}_{:,1}\| \, \|\tilde{\mathbf{J}}_{:,2}\|
}
\right).
\]

\begin{figure}
\centering
\includegraphics[width=0.8\linewidth]{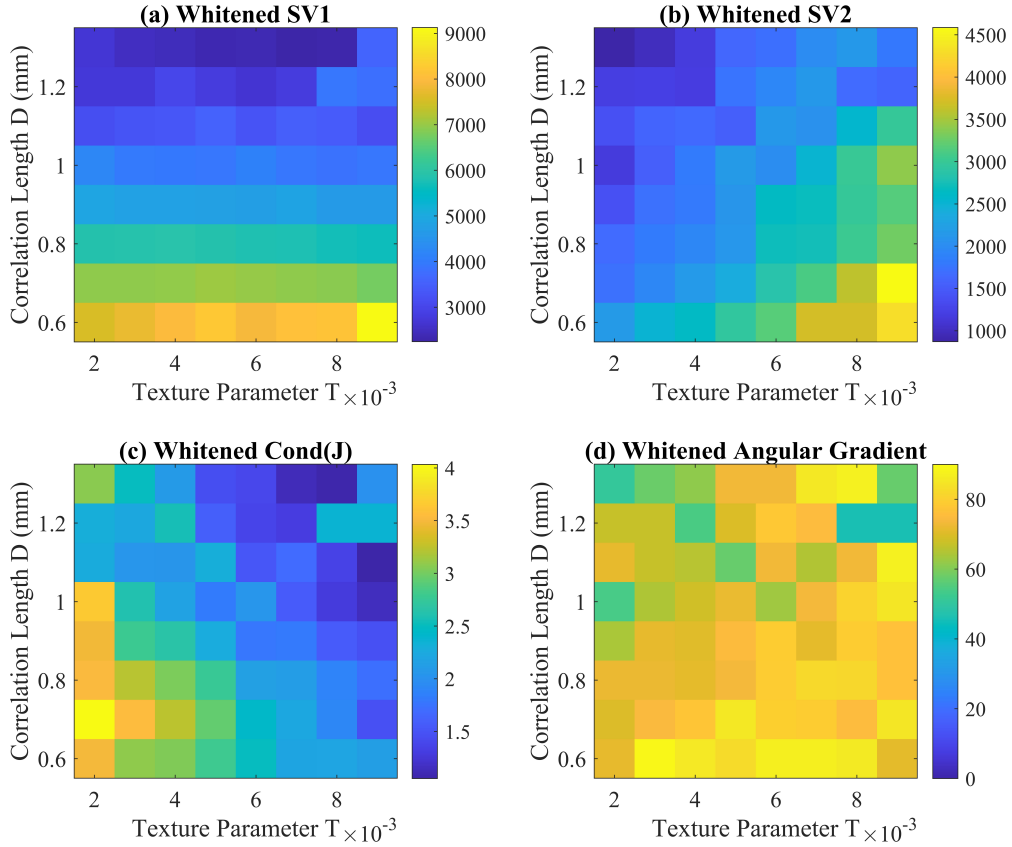}
\caption{
Whitened sensitivity metrics across the $(D,T)$ domain, incorporating intrinsic microstructure variability. (a)–(b) show the singular values of the whitened Jacobian, (c) the corresponding condition number, and (d) the angular separation between sensitivity directions. Whitening scales the Jacobian by the inverse covariance of the observables, so that sensitivity is measured relative to variability. This reflects practical identifiability and highlights regions where parameter influence is strong relative to stochastic microstructure fluctuations.}
\label{fig:GRF4}
\end{figure}

The variance-weighted Jacobian results are illustrated in Figure~\ref{fig:GRF4}. In contrast to the structural Jacobian, these quantities reflect practical identifiability by evaluating sensitivity relative to observable variability. While the forward map remains structurally full rank, the whitened metrics reveal strong anisotropy in effective sensitivity across the $(D,T)$ domain. In particular, the smallest singular value (SV2) is markedly reduced in regions of low $T$ and larger $D$, indicating that the weaker parameter direction becomes progressively less recoverable once stochastic variability is taken into account.

This behaviour reflects the interplay between sensitivity and scattering-induced variability. From Figure~\ref{fig:GRF2}(c)-(d), large $D$ and small $T$ values generated greater variability in the observables which in turn reduced the SV significantly in the area. Taking variability into account, the parameter-dependent signal is effectively masked by the intrinsic microstructural spread. As a result, practical identifiability is governed not by the magnitude of the Jacobian alone, but by how strongly the mean sensitivity rises above this variability floor.

The angular metric in Figure~\ref{fig:GRF4}(d) further shows that the sensitivity directions associated with $D$ and $T$ remain close to orthogonal over most of the parameter space. Hence, the variance-weighted sensitivity structure is not governed by directional collapse or parameter alignment. Instead, whitening reweighs the effective strength of the two directions according to their variability while preserving their geometric separation. This also explains why the whitened Jacobian shows a smaller SV1 and SV2 separation, and thus a more moderate condition number when compared to the raw Jacobian in Figure~\ref{fig:GRF3}.

\subsection{Practical Implications of Sensitivity Analysis}

\subsubsection{Practical Identifiability Map}

As each Jacobian-derived metric captures a different aspect of the local information geometry, it is not always straightforward to assess the overall identifiability of the system directly. A composite metric is therefore introduced to summarise both the strength and balance of the local information structure. The first term, $a$, represents the information floor, defined from the log-scaled minimum eigenvalue of the Fisher information matrix, and reflects the effective strength of the weakest parameter direction after variance normalisation. The second term represents a softened anisotropy measure, defined as
\begin{equation}
b = \sqrt{\frac{\lambda_{\min}}{\lambda_{\max}}},
\end{equation}
which captures the balance between parameter sensitivities and penalises highly elongated information geometries. These two components are normalised and combined through an arithmetic blend to yield the final practical identifiability index:
\begin{equation}
\mathrm{PI}(\boldsymbol{P}) = \frac{1}{2}\left(a + b\right),
\end{equation}
This index is not intended to be unique, but rather to provide a compact summary of the joint effects of information strength and directional balance. High values therefore correspond to regions where the weakest parameter direction remains informative and the local sensitivity geometry is relatively well balanced.

\begin{figure}
\centering
\includegraphics[width=0.47\linewidth]{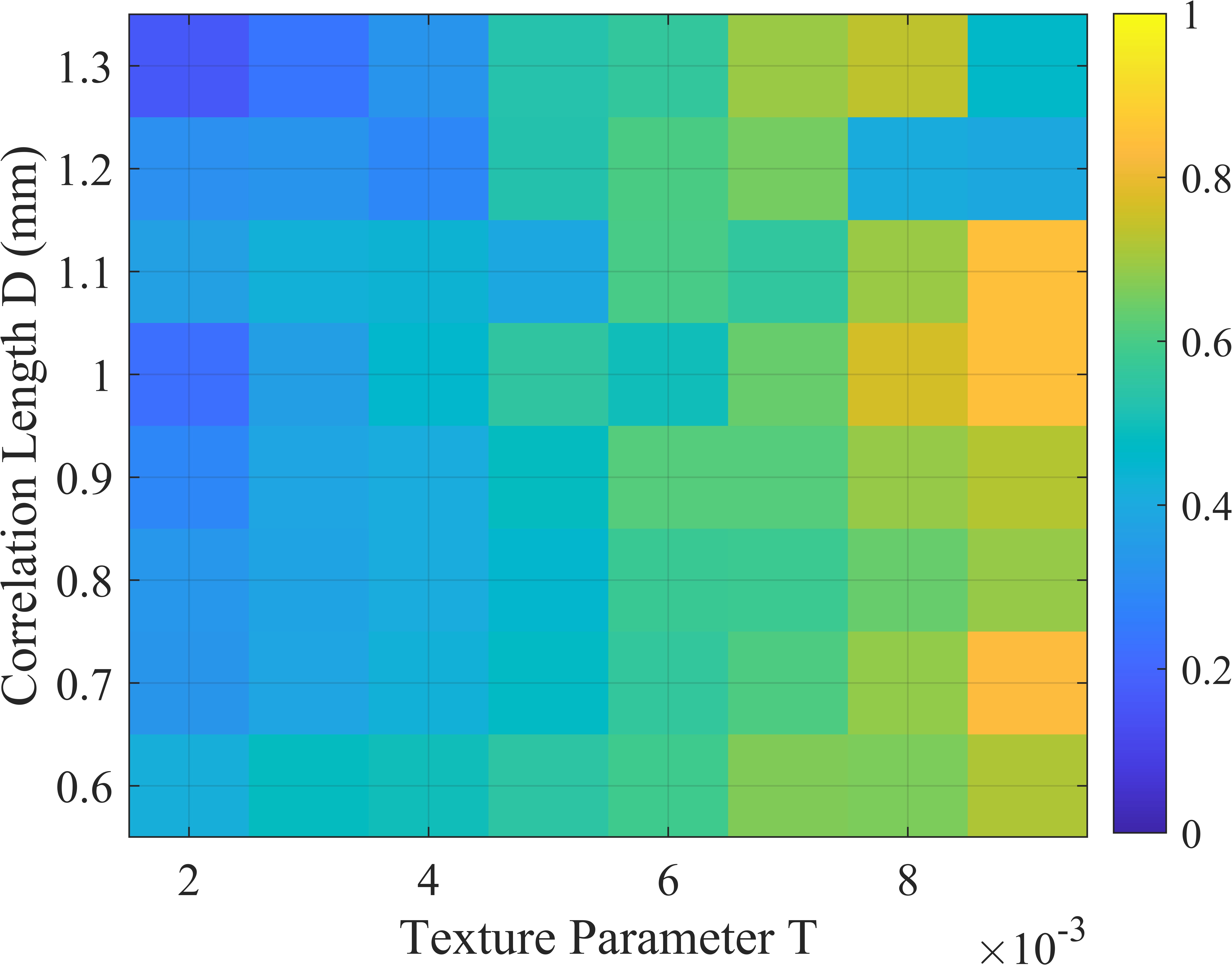}
\caption{Practical identifiability map over the $(D,T)$ parameter space based on a composite index combining the variance-weighted information floor and a softened anisotropy measure. Higher values indicate regions where the weakest parameter direction remains more informative and the local sensitivity geometry is more balanced, corresponding to more favourable inverse-problem conditioning.}
\label{fig:GRF6}
\end{figure}

The practical identifiability map in Figure~\ref{fig:GRF6} shows that the dominant control on PI is the texture-coherence parameter $T$, with PI increasing systematically with $T$. This indicates that $T$ primarily governs whether the weaker parameter direction remains practically resolvable after variance-weighting. At low $T$, the inverse problem is limited by insufficient stable information in the weakest direction; at higher $T$, that limitation is progressively relaxed and the local information geometry becomes more favourable for joint recovery of $(D,T)$. The effect of correlation length $D$ is more regime-dependent and acts mainly as a secondary modifier of this trend. PI does not vary monotonically with $D$, but instead shows modest variations in the domain. This suggests that $D$ influences practical identifiability primarily through how heterogeneity scale modulates the balance between variance-weighted sensitivity and scattering-induced variability. 

Practical identifiability therefore emerges from the coupled interaction between texture-coherence strength and heterogeneity scale, and is governed by the joint balance between information strength and directional conditioning rather than by any single parameter or metric alone. The proposed index provides a compact summary of this local information geometry, allowing favourable operating regimes to be identified and linked to inversion behaviour. It should therefore be interpreted as a relative diagnostic of inversion favourability, rather than as an absolute criterion for recoverability.

\subsubsection{Inversion of Surrogate Model}

To demonstrate how the sensitivity structure affects parameter recovery, inversion was carried out over the $(D,T)$ parameter space using a grid-based search method. The objective function was defined by the least-squares mismatch between an observed feature vector and the ensemble-mean surrogate prediction:
\begin{equation}
\mathcal{L}(\boldsymbol{P}) = 
\left\| \boldsymbol{\mu}_{\mathrm{obs}} - \boldsymbol{\mu}(\boldsymbol{P}) \right\|_2^2,
\end{equation}
where $\boldsymbol{\mu}_{\mathrm{obs}}$ denotes the observed mean feature vector. Locally, the shape of this objective is governed by the Jacobian $\mathbf{J}_{\mu}$ and the associated Fisher information, which determine how strongly and how independently perturbations in $(D,T)$ are reflected in the observables.

\begin{figure}
\centering
\includegraphics[width=0.9\linewidth]{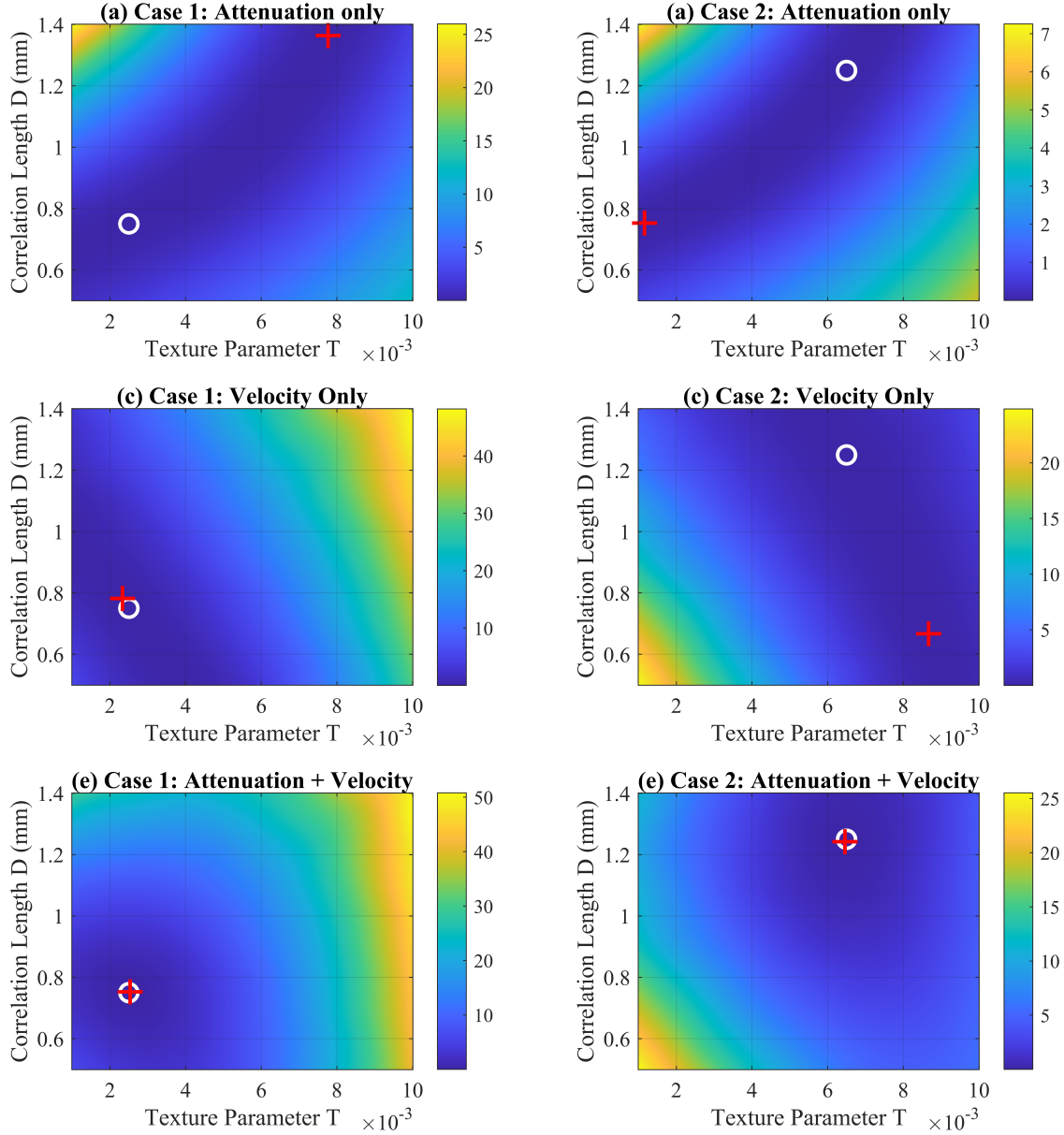}
\caption{Inversion landscapes for two representative parameter configurations using attenuation only, velocity only, and both observables for inversion. The colour map shows the least-squares objective $\mathcal{L}(D,T)$, with the white markers denoting the ground-truth and red markers the recovered solutions.}
\label{fig:GRF5}
\end{figure}

The inversion landscapes are compared in Figure~\ref{fig:GRF5} for two representative parameter configurations using attenuation only, velocity only, and both observables. When attenuation is used alone in (a)–(b), the objective surface has an elongated valley which denotes a weakly constrained parameter direction. A similar behaviour is seen in the velocity-only inversion (c)–(d) but with a different orientation. In both cases, the recovered solution is poorly localised and strays away from the ground-truth. This is consistent with the local sensitivity analysis, where a single observable provides only one dominant constraint and is therefore insufficient to resolve both parameters reliably. When attenuation and velocity are used together in (e)–(f), the objective surface becomes more compact, with a well-localised minimum at the ground-truth parameter location.

Overall, the inversion results confirm that accurate recovery requires complementary constraint across parameter directions: single observables yield weakly localised solutions, whereas the combined use of attenuation and velocity leads to substantially better recovery. Hence, inversion quality depends not only on the magnitude of sensitivity, but also on whether the chosen observables constrain complementary directions in the parameter space.

\section{Discussion}
\label{sec:discussion}

The results show that identifiability in ultrasonic microstructure characterisation is limited by different mechanisms across the canonical and surrogate models. In the canonical case, the main constraints arise from observable dimensionality and parameter coupling. In the surrogate case, the forward map remains structurally identifiable, but sensitivity is uneven across parameter directions and becomes more restrictive once stochastic variability is included. Inversion difficulty therefore depends not only on the reconstruction method, but on how the measurement encodes parameter changes into observable responses.

\subsection{From Canonical To Surrogate Identifiability}

In the canonical slab, the limitation is explicit: the observable set is too small to identify all parameters independently, and parameter coupling produces a near-null direction. This motivates the surrogate setup adopted here, where two surrogate descriptors and two observables are used to eliminate dimensional restriction. This enables the analysis to focus on anisotropy and variability rather than trivial rank deficiency. The local map from $(D,T)$ to the chosen observables therefore remains full rank, but the two parameter directions are not represented equally, since the relative effects of $D$ and $T$ are uneven and vary across the parameter space. The resulting inverse problem is therefore not limited by strict degeneracy, but by a regime-dependent imbalance in how strongly the two parameter directions are expressed in the observable space. This progression is summarised in Table~\ref{tab:ident_summary}, from rank-deficient canonical behaviour to full-rank but anisotropic surrogate sensitivity, and finally to variability-limited practical identifiability in the whitened setting.

\begin{table}[t]
\centering
\caption{Summary of identifiability regimes across modelling levels.}
\label{tab:ident_summary}
\begin{tabular}{lccc}
\hline
\textbf{Regime} & \textbf{Structural Identifiability} & \textbf{Practical Identifiability} & \textbf{Limiting Factor} \\
\hline
Canonical (1D) & Rank-deficient & Poor & Parameter coupling \\
Surrogate (2D) & Full rank & Anisotropic & Sensitivity imbalance \\
Surrogate (Variance-Weighted) & Full rank & Degraded & Stochastic variance \\
\hline
\end{tabular}
\end{table}

\subsection{Observability of Microstructure Descriptors}

Within the surrogate framework, attenuation and velocity carry different but complementary information about the underlying microstructure. Velocity is governed mainly by the effective elastic response of the medium and is more sensitive to the texture-coherence parameter $T$. Attenuation, by contrast, is more strongly influenced by distributed scattering and correlation length $D$, while remaining modulated by $T$. As both observables are not fully separable, it limits how reliably the coupled descriptors can be recovered. Hence, a strong signal alone does not guarantee a well-posed inverse problem if the corresponding sensitivity is weak, unevenly distributed, or suppressed by intrinsic variability. The inversion results are also consistent with this finding, where the objective landscape is elongated and weakly constrained when only one observable is used, while combining the complementary observables yield a localised minimum.

The inversion behaviour observed here should also be viewed in the context of real polycrystalline materials, where attenuation is influenced by more than a single characteristic grain size. In particular, grain-size distributions can modify the attenuation response in ways that are not captured by a single scalar descriptor, so distinct microstructural states may remain difficult to separate using a limited observable set. Similar considerations have been reported in both theoretical and numerical studies of polycrystals with distributed grain sizes \cite{Arguelles2017,Yeoh2025}. In such cases, attenuation may be non-unique and velocity may provide limited discrimination when texture states are similar, increasing the value of additional observables such as backscatter \cite{Liu2019}. The present surrogate inversions should therefore be interpreted not as literal reconstructions of real microstructures, but as controlled examples of a broader identifiability issue that is also relevant in practical materials characterisation.

\subsection{Implications for Measurement Design and Inference}

These results suggest that ultrasonic measurement design should be guided by the local information geometry of the inverse problem rather than by signal amplitude or any single sensitivity metric alone \cite{Huan2024}. Better recoverability is expected when the observable set provides sufficiently strong, balanced, and variability-robust constraint across parameter directions. In practice, this favours multi-observable, multi-frequency, or multi-configuration interrogation strategies, as well as richer observables such as angular scattering information, backscatter measurements, or spatially resolved wavefields.

An analogous example is in ultrasonic imaging, where full matrix capture (FMC) combined with total focusing method (TFM) improves image fidelity by using a richer set of transmit--receive information channels \cite{Holmes2005,Wilcox2007}. Although the present problem concerns parameter inference rather than image formation, the same idea applies here: recoverability improves when the measurement provides multiple complementary constraints on the unknowns.

The same argument also applies to learning-based inference. In both the canonical and surrogate settings, the recoverable content of the measurement is still limited by the structure of the forward map. If sensitivity to a parameter is weak or strongly suppressed relative to variability, then increasing model complexity or training data alone will not reliably recover that parameter \cite{Tachella2022}. Data-driven methods may improve robustness within the identifiable subspace, but they cannot reconstruct information that is only weakly expressed in the measurement itself \cite{Arridge2019}. Viewed in this way, the feature-level Fisher framework is useful not only for analysing inversion difficulty after the fact, but also for comparing measurement strategies in advance and identifying which observable sets are likely to provide stronger parameter resolution.

\subsection{Limitations and Future Work}

The present study is subject to several limitations, primarily associated with the surrogate microstructural representation and the restricted observable framework.

First, the surrogate microstructure model is based on stochastic Gaussian random fields (GRFs) parameterised by correlation length $D$ and texture-coherence parameter $T$. Although this provides a controlled setting for isolating the effects of heterogeneity scale, scattering, and variability, it does not explicitly represent discrete grain morphologies, grain boundary networks, or full crystallographic detail as found in real polycrystals. The descriptors used here should therefore be interpreted as effective surrogate parameters rather than direct metallographic quantities.

Second, the analysis is restricted to two-dimensional microstructures. Although the main identifiability trends are expected to persist in three dimensions, the additional degrees of freedom in three dimensions can alter the scattering behaviour and the coupling between descriptors and observables \cite{VanPamel2016j}. Hence, the extension to explicit three-dimensional grain structures is necessary to assess these effects quantitatively.

Third, in this study with surrogate microstructural model, only two descriptors were evaluated, the correlation length $D$ and texture-coherence $T$ parameter. Actual microstructures contain other important characteristics such as grain morphology, grain-size distribution, and phases. These descriptors need to be incorporated to extend the identifiability analysis to a wider range of engineering materials.

Finally, the paper covers only attenuation- and velocity-based scalar features. While this keeps the analysis interpretable and closely tied to experimentally accessible measurements, it also restricts the observable space and exclude information strands carried by other wave modalities. Hence, additional observables such as backscatter amplitude, angular scattering information, and mode-converted components can be considered as they may provide stronger independent constraints.

The identifiability limits and trends reported here should therefore be interpreted within the present modelling and observable framework, rather than as universal limits for ultrasonic materials characterisation. Future work will extend the analysis to more realistic three-dimensional microstructures with additional descriptors, richer observables, and experimental validation, in order to assess how these identifiability trends carry over to practical engineering material systems.

\section{Conclusions}
\label{sec:conclusions}

This work examined identifiability limits in ultrasonic microstructure characterisation using a combined analytical and numerical framework. In the canonical pulse--echo model, closed-form sensitivities were used to establish the baseline information limits under idealised conditions. The results showed that recoverability is restricted by three related effects: dimensional limitation, parameter coupling, and impedance-dependent saturation. In particular, the conditioning of the inverse problem changes strongly with impedance contrast, while sensitivity to impedance weakens as the reflection coefficients approach saturation. These effects arise from the structure of the forward map itself, rather than from the inversion method used to interrogate it.

The framework was then extended to statistically generated surrogate microstructures. For the two-parameter surrogate model considered here, the map from correlation length $D$ and texture-coherence parameter $T$ to attenuation and velocity remained structurally full rank over the explored domain. The main limitation was therefore no longer the number of observables, but how unevenly sensitivity was distributed across the two parameter directions. Both $D$ and $T$ influence the chosen observables with differing strength, with the weaker direction being difficult to resolve in parts of the parameter space. In this setting, anisotropy arises primarily from imbalance in sensitivity magnitude rather than from directional collinearity.

Once stochastic variability across microstructure realisations is included, the problem becomes more restrictive at the level of practical recovery. Although the two parameter directions remain largely distinct, their recoverability depends on whether the corresponding sensitivities remain strong relative to the variability floor. The variance-weighted Fisher analysis showed that the weakest direction can be substantially suppressed once this variability is accounted for, reducing practical resolution even though the mean forward map remains structurally identifiable. This is further demonstrated with the inversion examples, whereby single observables produced elongated objective landscapes with weak constraints, whereas the combined inversion gave a more localised minimum. This behaviour reflects the underlying information geometry, where recovery improves when the observable set constrains complementary directions in parameter space.

In conclusion, the canonical and surrogate analyses identify three distinct sources of difficulty: dimensional restriction and parameter coupling in the canonical model, anisotropic but full-rank sensitivity in the surrogate model, and variance-induced suppression at the level of practical recovery. The framework developed here provides a systematic way to distinguish between these regimes and to assess whether a limitation arises from structural non-identifiability, poor conditioning, or variability-limited resolution. Within the modelling and observable framework considered here, these results show that identifiability analysis is valuable not only for interpreting inversion performance, but also for guiding observable selection and measurement design in ultrasonic microstructure characterisation. 

\bibliographystyle{unsrtnat}

\bibliography{references}

@article{shukla2022physics,
  title   = {A Physics-Informed Neural Network for Quantifying the Microstructural Properties of Polycrystalline Nickel Using Ultrasound Data: A Promising Approach for Solving Inverse Problems},
  author  = {Shukla, Khemraj and Jagtap, Ameya D. and Blackshire, James L. and Sparkman, Daniel and Karniadakis, George Em},
  journal = {IEEE Signal Processing Magazine},
  volume  = {39},
  number  = {1},
  pages   = {68--77},
  year    = {2022},
  doi     = {10.1109/MSP.2021.3118904}
}

@article{patel2025developing,
  title   = {Developing neural networks to rapidly map crystallographic orientation using laser ultrasound measurements},
  author  = {Patel, Rikesh and Li, Wenqi and Smith, Richard J. and Clark, Matt},
  journal = {Scripta Materialia},
  volume  = {256},
  pages   = {116415},
  year    = {2025},
  doi     = {10.1016/j.scriptamat.2024.116415}
}

@article{singh2022deep,
  title   = {Deep Learning Based Inversion of Locally Anisotropic Weld Properties from Ultrasonic Array Data},
  author  = {Singh, Jonathan and Tant, Katherine and Mulholland, Anthony and MacLeod, Charles},
  journal = {Applied Sciences},
  volume  = {12},
  number  = {2},
  pages   = {532},
  year    = {2022},
  doi     = {10.3390/app12020532}
}

@article{liu2022autonomous,
  title   = {Autonomous characterization of grain size distribution using nonlinear Lamb waves based on deep learning},
  author  = {Liu, Lishuai and Wu, Peng and Xiang, Yanxun and Xuan, Fu-Zhen},
  journal = {The Journal of the Acoustical Society of America},
  volume  = {152},
  number  = {3},
  pages   = {1913--1921},
  year    = {2022},
  doi     = {10.1121/10.0014289}
}

@article{vela2026characterization,
  title   = {Characterization of microtexture in Ti-6Al-4 V using ultrasonic polar forward scattering},
  author  = {Vela, Ramon, III and Matz, Nathanial J. and Hassan, Waled and Turner, Joseph A.},
  journal = {Ultrasonics},
  volume  = {165},
  pages   = {108113},
  year    = {2026},
  doi     = {10.1016/j.ultras.2026.108113}
}

@article{liu2025combined,
  title   = {Combined effect of frequency, grain size, and sound path on phased array ultrasonic spectrum and attenuation},
  author  = {Liu, Yu and Tian, Qiang and Zhang, Shuzeng and Guan, Xuefei},
  journal = {Ultrasonics},
  volume  = {149},
  pages   = {107593},
  year    = {2025},
  doi     = {10.1016/j.ultras.2025.107593}
}

@article{liu2023spatial,
  title   = {Spatial and directional characterization of wire and arc additive manufactured aluminum alloy using phased array ultrasonic backscattering method},
  author  = {Liu, Yu and Wang, Xinyan and Oliveira, J. P. and He, Jingjing and Guan, Xuefei},
  journal = {Ultrasonics},
  volume  = {132},
  pages   = {107024},
  year    = {2023},
  doi     = {10.1016/j.ultras.2023.107024}
}

@article{kalkowski2021grazing,
  title   = {How does grazing incidence ultrasonic microscopy work? A study based on grain-scale numerical simulations},
  author  = {Kalkowski, Micha{\l} K. and Lowe, Michael J. S. and Barth, Martin and Rjelka, Marek and K{\"o}hler, Bernd},
  journal = {Ultrasonics},
  volume  = {114},
  pages   = {106387},
  year    = {2021},
  doi     = {10.1016/j.ultras.2021.106387}
}

@article{victoriagiraldo2025ultrasonic,
  title   = {Ultrasonic scattering in polycrystalline materials with elongated grains: A comparative 3D and 2D theoretical and numerical analysis},
  author  = {Victoria-Giraldo, Juan Camilo and Tie, Bing and Laurent, J{\'e}r{\^o}me and Lh{\'e}mery, Alain and Solas, Denis},
  journal = {Ultrasonics},
  volume  = {152},
  pages   = {107642},
  year    = {2025},
  doi     = {10.1016/j.ultras.2025.107642}
}

@article{ruiz2024ultrasonic,
  title   = {Ultrasonic characterization of microstructural changes due to static recrystallization and grain growth in Inconel 625 and Inconel 718 superalloys},
  author  = {Ruiz, Alberto and Rodr{\'i}guez, Vania M. and Granados Becerra, Heriberto and Kim, Jin-Yeon},
  journal = {Ultrasonics},
  volume  = {142},
  pages   = {107383},
  year    = {2024},
  doi     = {10.1016/j.ultras.2024.107383}
}

@article{li2024bs,
  title   = {Ultrasonic backscattering model for Rayleigh waves in polycrystals with Born and independent scattering approximations},
  author  = {Li, Shan and Huang, Ming and Song, Yongfeng and Lan, Bo and Li, Xiongbing},
  journal = {Ultrasonics},
  volume  = {140},
  pages   = {107297},
  year    = {2024},
  doi     = {10.1016/j.ultras.2024.107297}
}

@article{liu2025ultrasonic,
  title   = {Ultrasonic backscattering method for characterizing the non-uniform microstructure of polycrystals},
  author  = {Liu, Bohan and Huang, Ming and Zhang, Dehan and Yu, Xudong},
  journal = {International Journal of Mechanical Sciences},
  volume  = {306},
  pages   = {110831},
  year    = {2025},
  doi     = {10.1016/j.ijmecsci.2025.110831}
}

@article{grabec2022synthetic,
  title   = {Surface acoustic wave attenuation in polycrystals: Numerical modeling using a statistical digital twin of an actual sample},
  author  = {Grabec, Tom{\'a}{\v{s}} and Veres, Istv{\'a}n A. and Ryzy, Martin},
  journal = {Ultrasonics},
  volume  = {119},
  pages   = {106585},
  year    = {2022},
  doi     = {10.1016/j.ultras.2021.106585}
}

@article{liu2022shape,
  title   = {Shape and size evaluations of elongated grains using phased array ultrasound and directional backscattering method},
  author  = {Liu, Yu and Tian, Qiang and Yu, Ping and He, Jingjing and Guan, Xuefei},
  journal = {NDT \& E International},
  volume  = {129},
  pages   = {102634},
  year    = {2022},
  doi     = {10.1016/j.ndteint.2022.102634}
}

@article{victoriagiraldo2026texture,
  title   = {Propagation and scattering of ultrasonic waves in macroscopically anisotropic polycrystalline materials with fiber texture},
  author  = {Victoria-Giraldo, Juan Camilo and Solas, Denis and Laurent, J{\'e}r{\^o}me and Lh{\'e}mery, Alain and Tie, Bing},
  journal = {Ultrasonics},
  volume  = {164},
  pages   = {108026},
  year    = {2026},
  doi     = {10.1016/j.ultras.2026.108026}
}

@article{dorval2025numerical,
  title   = {Numerical estimation of ultrasonic phase velocity and attenuation for longitudinal and shear waves in polycrystalline materials},
  author  = {Dorval, Vincent and Leymarie, Nicolas and Imperiale, Alexandre and Demaldent, Edouard and Lhuillier, Pierre-Emile},
  journal = {Ultrasonics},
  volume  = {148},
  pages   = {107517},
  year    = {2025},
  doi     = {10.1016/j.ultras.2024.107517}
}

@article{li2025backscatter,
  title   = {Effect of microstructure and texture gradient on the backscattered ultrasound amplitude of Ti-6Al-4V bars for aeroengine blade},
  author  = {Li, Lei and Ying, Yang and Zhou, Min and Cao, Zuhan and Guo, Dizi and Yang, Haiying and Ding, Li and Han, Feixiao and Vincent, Ji},
  journal = {Journal of Alloys and Compounds},
  volume  = {1010},
  pages   = {177604},
  year    = {2025},
  doi     = {10.1016/j.jallcom.2024.177604}
}

@article{renaud2021multiparameter,
  title   = {Multi-parameter optimization of attenuation data for characterizing grain size distributions and application to bimodal microstructures},
  author  = {Renaud, Adrien and Tie, Bing and Mouronval, Anne-Sophie and Schmitt, Jean-Hubert},
  journal = {Ultrasonics},
  volume  = {115},
  pages   = {106425},
  year    = {2021},
  doi     = {10.1016/j.ultras.2021.106425}
}

@article{dryburgh2020velmap,
  title   = {Determining the crystallographic orientation of hexagonal crystal structure materials with surface acoustic wave velocity measurements},
  author  = {Dryburgh, Paul and Smith, Richard J. and Marrow, Paul and Lain{\'e}, Steven J. and Sharples, Steve D. and Clark, Matt and Li, Wenqi},
  journal = {Ultrasonics},
  volume  = {108},
  pages   = {106171},
  year    = {2020},
  doi     = {10.1016/j.ultras.2020.106171}
}

@article{guo2026decoupling,
  title   = {Decoupling analysis of ultrasonic scattering characteristics in porous polycrystalline materials using phase field and finite element methods},
  author  = {Guo, Zixin and Song, Yongfeng and Li, Xiongbing},
  journal = {Ultrasonics},
  volume  = {159},
  pages   = {107864},
  year    = {2026},
  doi     = {10.1016/j.ultras.2025.107864}
}

@article{Sun2024VelocityMap,
  author  = {Sun, Zeqing and Wu, Shangzi and Saini, Abhishek and Fan, Zheng},
  title   = {Characterizing polycrystalline microstructures by reconstructing grain boundaries and velocity map from ultrasonic surface wave field},
  journal = {Ultrasonics},
  volume  = {137},
  pages   = {107175},
  year    = {2024},
  doi     = {10.1016/j.ultras.2023.107175}
}

@article{He2024CodaWaveGrainSize,
  author  = {He, Jingjing and Gao, Chenjun and Wang, Xun and Yang, Jinsong and Tian, Qiang and Guan, Xuefei},
  title   = {Average grain size evaluation using scattering-induced attenuation of coda waves},
  journal = {Ultrasonics},
  volume  = {141},
  pages   = {107334},
  year    = {2024},
  doi     = {10.1016/j.ultras.2024.107334}
}

@article{Sheng2026BroadGrainSize,
  author  = {Sheng, Ningyue and Khazaie, Shahram},
  title   = {A new correlation model for ultrasonic attenuation in polycrystals with broad grain size distributions},
  journal = {Ultrasonics},
  volume  = {160},
  pages   = {107924},
  year    = {2026},
  doi     = {10.1016/j.ultras.2025.107924}
}

@article{Gong2026,
  author  = {Gong, Shaojie and Guo, Shifeng and Xiong, Yi and Zhou, Shiyuan and Cui, Fangsen and Liu, Menglong},
  title   = {Wave propagation in highly anisotropic polycrystals: A numerical perspective from an unstructured-mesh-based high-order finite element method},
  journal = {Ultrasonics},
  volume  = {159},
  pages   = {107882},
  year    = {2026},
  doi     = {10.1016/j.ultras.2025.107882}
}

@article{Liu2025_Duplex,
  author  = {Liu, Zenghua and Li, Jinlong and Zheng, Yang and He, Cunfu},
  title   = {Ultrasonic backscattering model of lamellar duplex phase microstructures in polycrystalline materials},
  journal = {Ultrasonics},
  volume  = {149},
  pages   = {107581},
  year    = {2025},
  doi     = {10.1016/j.ultras.2025.107581}
}

@article{Arridge2019,
  author  = {Arridge, Simon and Maass, Peter and \"Oktem, Ozan and Sch\"onlieb, Carola-Bibiane},
  title   = {Solving inverse problems using data-driven models},
  journal = {Acta Numerica},
  volume  = {28},
  pages   = {1--174},
  year    = {2019},
  doi     = {10.1017/S0962492919000059}
}

@inproceedings{Tachella2022,
  author    = {Tachella, Julian and Hurault, Samuel and Davies, Mike},
  title     = {Unsupervised Learning from Incomplete Measurements for Inverse Problems},
  booktitle = {Advances in Neural Information Processing Systems},
  volume    = {35},
  pages     = {4981--4993},
  year      = {2022}
}

@article{Huan2024,
  author  = {Huan, Xun and Marzouk, Youssef M.},
  title   = {Optimal experimental design: Formulations and computations},
  journal = {Acta Numerica},
  volume  = {33},
  pages   = {1--140},
  year    = {2024},
  doi     = {10.1017/S0962492924000013}
}

@article{Wilcox2007,
  author  = {Wilcox, Paul D. and Holmes, Caroline and Drinkwater, Bruce W.},
  title   = {Advanced reflector characterization with ultrasonic phased arrays in NDE applications},
  journal = {IEEE Transactions on Ultrasonics, Ferroelectrics, and Frequency Control},
  volume  = {54},
  number  = {8},
  pages   = {1541--1550},
  year    = {2007},
  doi     = {10.1109/TUFFC.2007.424}
}

@book{Li2017,
  title     = {Numerical Solution of Differential Equations: Introduction to Finite Difference and Finite Element Methods},
  author    = {Li, Zhilin and Qiao, Zhonghua and Tang, Tao},
  publisher = {Cambridge University Press},
  year      = {2017},
  doi       = {10.1017/9781316671528}
}

@article{Schubert2004,
  title = {Numerical time-domain modeling of linear and nonlinear ultrasonic wave propagation using finite integration techniques––theory and applications},
  author = {Frank Schubert},
  journal = {Ultrasonics},
  volume = {42},
  number = {1-9},
  pages = {221--229},
  year = {2004},
  issn = {0041-624X},
  doi = {10.1016/j.ultras.2004.01.013}
}

@article{Alford1974,
  author = {Alford, R. M. and Kelly, K. R. and Whitmore, D. M.},
  title = {Accuracy of finite-difference modeling of the acoustic wave equation},
  journal = {Geophysics},
  volume = {39},
  number = {6},
  pages = {834--842},
  year = {1974},
  doi = {10.1190/1.1440470}
}

@phdthesis{Zakaria2003,
  title  = {Numerical Modelling of Wave Propagation Using Higher Order Finite-Difference Formulas},
  author = {Zakaria, Ahmad},
  school = {Curtin University of Technology},
  year   = {2003},
  month  = {May},
  type   = {PhD Thesis},
  url    = {https://espace.curtin.edu.au/handle/20.500.11937/190}
}

@article{Liu2019a,
  author = {Liu, Yang and Li, Jingfa and Sun, Shuyu and Yu, Bo},
  title = {Advances in Gaussian random field generation: A review},
  journal = {IEEE Access},
  volume = {7},
  pages = {153123--153139},
  year = {2019},
  doi = {10.1109/ACCESS.2019.2948616}
}

@article{Bellman1970,
  title = {On structural identifiability},
  author = {Bellman, R. and {\AA}str{\"o}m, K. J.},
  journal = {Mathematical Biosciences},
  volume = {7},
  number = {3-4},
  pages = {329--339},
  year = {1970},
  issn = {0025-5564},
  doi = {10.1016/0025-5564(70)90132-X}
}

@article{AnstettCollin2020,
  title = {A priori identifiability: {An} overview on definitions and approaches},
  author = {Anstett-Collin, F. and Denis-Vidal, L. and Mill{\'e}rioux, G.},
  journal = {Annual Reviews in Control},
  volume = {50},
  pages = {139--149},
  year = {2020},
  issn = {1367-5788},
  doi = {10.1016/j.arcontrol.2020.10.006}
}

@article{Lam2022,
  title = {Practical identifiability of parametrised models: {A} review of benefits and limitations of various approaches},
  author = {Lam, Nicholas N. and Docherty, Paul D. and Murray, Rua},
  journal = {Mathematics and Computers in Simulation},
  volume = {199},
  pages = {202--216},
  year = {2022},
  issn = {0378-4754},
  doi = {10.1016/j.matcom.2022.03.020}
}

@article{Guillaume2019,
  title = {Introductory overview of identifiability analysis: {A} guide to evaluating whether you have the right type of data for your modeling purpose},
  author = {Guillaume, Joseph H. A. and Jakeman, John D. and Marsili-Libelli, Stefano and Asher, Michael and Brunner, Philip and Croke, Barry and Hill, Mary C. and Jakeman, Anthony J. and Keesman, Karel J. and Razavi, Saman and Stigter, Johannes D.},
  journal = {Environmental Modelling \& Software},
  volume = {119},
  pages = {418--432},
  year = {2019},
  issn = {1364-8152},
  doi = {10.1016/j.envsoft.2019.07.007}
}

@article{Cobelli1980,
  title = {Parameter and structural identifiability concepts and ambiguities: a critical review and analysis},
  author = {Cobelli, C. and DiStefano, III, J. J.},
  journal = {American Journal of Physiology-Regulatory, Integrative and Comparative Physiology},
  volume = {239},
  number = {1},
  pages = {R7--R24},
  year = {1980},
  doi = {10.1152/ajpregu.1980.239.1.R7}
}

@book{Rose2014,
author = {Rose, Joseph L.},
booktitle = {Ultrasonic Guided Waves in Solid Media},
chapter = {4},
pages = {53 -- 66},
title = {{Reflection and Refraction}},
publisher = {Cambridge University Press},
year = {2014}
}

@article{Viana2024,
  author  = {Viana, M. C. A. and Pereira, P. and Buenos, A. A. and Santos, A. A.},
  title   = {Identifying Grain Size in ASTM A36 Steel Using Ultrasonic Backscattered Signals and Machine Learning},
  journal = {NDT \& E International},
  year    = {2024},
  note    = {In press}
}

@article{Wang2025,
  author  = {Wang, Wei and Zhang, Jie and Wilcox, Paul D.},
  title   = {Metallic material microstructure grain size measurements from backscattered ultrasonic array data using full matrix capture},
  journal = {NDT \& E International},
  year    = {2025},
  volume  = {149},
  pages   = {103251},
  doi     = {10.1016/j.ndteint.2024.103251},
  publisher = {Elsevier}
}

@article{Yeoh2025,
  title   = {Representative microstructures for two-dimensional computational studies of ultrasonic wave propagation in titanium alloys},
  author  = {Yeoh, W. Y. and Lan, B. and Lowe, M. J. S.},
  journal = {Proceedings of the Royal Society A: Mathematical, Physical and Engineering Sciences},
  year    = {2025},
  volume  = {481},
  number  = {2323},
  pages   = {20240776},
  doi     = {10.1098/rspa.2024.0776}
}

@article{Lobkis2012b,
author = {Lobkis, O. I. and Yang, L. and Li, J. and Rokhlin, S. I.},
doi = {10.1016/j.ultras.2011.12.002},
issn = {0041624X},
journal = {Ultrasonics},
number = {6},
pages = {694--705},
title = {{Ultrasonic backscattering in polycrystals with elongated single phase and duplex microstructures}},
volume = {52},
year = {2012}
}

@article{VanPamel2018a,
author = {{Van Pamel}, A. and Sha, G. and Lowe, M. J. S. and Rokhlin, S. I.},
doi = {10.1121/1.5031008},
isbn = {9683631088},
issn = {0001-4966},
journal = {The Journal of the Acoustical Society of America},
number = {4},
pages = {2394--2408},
title = {{Numerical and analytic modelling of elastodynamic scattering within polycrystalline materials}},
volume = {143},
year = {2018}
}

@inproceedings{Kube2015,
  author = {Kube, Christopher M. and Turner, Joseph A.},
  title = {{Voigt, Reuss, Hill, and Self-Consistent Techniques for Modeling Ultrasonic Scattering}},
  booktitle = {41st Annual Review of Progress in Quantitative Nondestructive Evaluation},
  volume = {1650},
  pages = {926--934},
  year = {2015},
  doi = {10.1063/1.4914698}
}

@article{Holmes2005,
author = {Holmes, Caroline and Drinkwater, Bruce W. and Wilcox, Paul D.},
doi = {10.1016/j.ndteint.2005.04.002},
issn = {09638695},
journal = {NDT and E International},
number = {8},
pages = {701--711},
title = {{Post-processing of the full matrix of ultrasonic transmit-receive array data for non-destructive evaluation}},
volume = {38},
year = {2005}
}

@article{Quey2011,
author = {Quey, R. and Dawson, P. R. and Barbe, F.},
doi = {10.1016/j.cma.2011.01.002},
issn = {00457825},
journal = {Computer Methods in Applied Mechanics and Engineering},
number = {17-20},
pages = {1729--1745},
publisher = {Elsevier B.V.},
title = {{Large-scale 3D random polycrystals for the finite element method: Generation, meshing and remeshing}},
volume = {200},
year = {2011}
}

@article{Bo2018,
author = {Lan, Bo and {Ben Britton}, T. and Jun, Tea-Sung and Gan, Weimin and Hofmann, Michael and Dunne, Fionn P.E. and Lowe, Michael J.S.},
doi = {10.1016/j.actamat.2018.08.037},
issn = {13596454},
journal = {Acta Materialia},
pages = {384--394},
publisher = {Elsevier Ltd},
title = {{Direct volumetric measurement of crystallographic texture using acoustic waves}},
volume = {159},
year = {2018}
}

@inproceedings{Margetan1993a,
author = {Margetan, F J and Thompson, R B},
booktitle = {Review of Progress in Quantitative Nondestructive Evaluation},
pages = {1735--1742},
title = {{Modeling Ultrasonic Microstructural Noise in Titanium Alloys}},
volume = {12},
year = {1993}
}

@article{Huthwaite2014,
author = {Huthwaite, Peter},
doi = {10.1016/j.jcp.2013.10.017},
isbn = {0021-9991},
issn = {00219991},
journal = {Journal of Computational Physics},
pages = {687--707},
publisher = {Elsevier Inc.},
title = {{Accelerated finite element elastodynamic simulations using the GPU}},
volume = {257},
year = {2014}
}

@article{Liu2019,
author = {Liu, Yuan and {Van Pamel}, Anton and Nagy, Peter B. and Cawley, Peter},
doi = {10.1121/1.5094783},
issn = {0001-4966},
journal = {The Journal of the Acoustical Society of America},
number = {3},
pages = {1584--1595},
title = {{Investigation of ultrasonic backscatter using three-dimensional finite element simulations}},
volume = {145},
year = {2019}
}

@article{Li2014a,
  title   = {Effect of texture and grain shape on ultrasonic backscattering in polycrystals},
  author  = {Li, J. and Yang, L. and Rokhlin, S. I.},
  journal = {Ultrasonics},
  volume  = {54},
  number  = {7},
  pages   = {1789--1803},
  year    = {2014},
  doi     = {10.1016/j.ultras.2014.02.020}
}

@article{Arguelles2016b,
author = {Arguelles, Andrea P. and Kube, Christopher M. and Hu, Ping and Turner, Joseph A.},
doi = {10.1121/1.4962161},
issn = {0001-4966},
journal = {The Journal of the Acoustical Society of America},
number = {3},
pages = {1570--1580},
title = {{Mode-converted ultrasonic scattering in polycrystals with elongated grains}},
volume = {140},
year = {2016}
}

@article{Lan2014a,
author = {Lan, B and Lowe, M. J. S. and Dunne, F. P. E.},
doi = {10.1016/j.actamat.2013.10.012},
issn = {1359-6454},
journal = {Acta Materialia},
pages = {107--122},
publisher = {Acta Materialia Inc.},
title = {{Experimental and computational studies of ultrasound wave propagation in hexagonal close-packed polycrystals for texture detection}},
volume = {63},
year = {2014}
}

@article{Weaver1990,
author = {Weaver, Richard L.},
doi = {10.1007/978-3-662-44324-8_1289},
journal = {Journal of the Mechanics and Physics of Solids},
number = {1},
pages = {55--86},
title = {{Diffusivity of Ultrasound in Polycrystals}},
volume = {38},
year = {1990}
}

@article{Lan2018,
author = {Lan, Bo and Carpenter, Michael A and Gan, Weimin and Hofmann, Michael and Dunne, Fionn P E and Lowe, Michael J S},
doi = {10.1016/j.scriptamat.2018.07.029},
issn = {1359-6462},
journal = {Scripta Materialia},
pages = {44--48},
publisher = {Elsevier Ltd.},
title = {{Rapid measurement of volumetric texture using resonant ultrasound spectroscopy}},
url = {https://doi.org/10.1016/j.scriptamat.2018.07.029},
volume = {157},
year = {2018}
}

@article{VanPamel2017a,
author = {{Van Pamel}, A. and Sha, G. and Rokhlin, S. I. and Lowe, M. J. S.},
doi = {10.1098/rspa.2016.0738},
isbn = {9781538633830},
issn = {1364-5021},
journal = {Proceedings of the Royal Society A: Mathematical, Physical and Engineering Science},
number = {2197},
title = {{Finite-element modelling of elastic wave propagation and scattering within heterogeneous media}},
volume = {473},
year = {2017}
}

@article{Stanke1984a,
author = {Stanke, Fred E. and Kino, G. S.},
doi = {10.1121/1.390577},
isbn = {0306651750},
issn = {0001-4966},
journal = {The Journal of the Acoustical Society of America},
number = {3},
pages = {665--681},
pmid = {26327816},
title = {{A unified theory for elastic wave propagation in polycrystalline materials}},
volume = {75},
year = {1984}
}

@article{Bai2020,
author = {Bai, X and Tie, B and Schmitt, J and Aubry, D},
doi = {10.1016/j.ultras.2019.105980},
issn = {0041-624X},
journal = {Ultrasonics},
number = {December 2018},
pages = {105980},
publisher = {Elsevier},
title = {{Comparison of ultrasonic attenuation within two- and three-dimensional polycrystalline media}},
volume = {100},
year = {2020}
}

@article{VanPamel2015a,
author = {{Van Pamel}, Anton and Brett, Colin R. and Huthwaite, Peter and Lowe, Michael J. S.},
doi = {10.1121/1.4931445},
isbn = {0001-4966},
issn = {0001-4966},
journal = {The Journal of the Acoustical Society of America},
number = {4},
pages = {2326--2336},
pmid = {26520313},
title = {{Finite element modelling of elastic wave scattering within a polycrystalline material in two and three dimensions}},
volume = {138},
year = {2015}
}

@article{Arguelles2017,
author = {Arguelles, Andrea P and Turner, Joseph A},
doi = {10.1121/1.4984290},
journal = {The Journal of the Acoustical Society of America},
number = {6},
pages = {4347 -- 4353},
title = {{Ultrasonic attenuation of polycrystalline materials with a distribution of grain sizes.}},
volume = {141},
year = {2017}
}

@article{VanPamel2016j,
author = {{Van Pamel}, Anton and Huthwaite, Peter and Brett, Colin R. and Lowe, Michael J.S.},
doi = {10.1016/j.ndteint.2016.02.004},
issn = {09638695},
journal = {NDT and E International},
pages = {9--19},
publisher = {Elsevier},
title = {{Numerical simulations of ultrasonic array imaging of highly scattering materials}},
url = {http://dx.doi.org/10.1016/j.ndteint.2016.02.004},
volume = {81},
year = {2016}
}

@book{Rose1999a,
author = {Rose, Joseph L.},
booktitle = {Ultrasonic Waves in Solid Media},
publisher = {Cambridge University Press},
chapter = {3},
pages = {24 -- 39},
title = {{Unbounded Isotropic and Anisotropic Media}},
year = {1999}
}

@article{Margetan1994,
author = {Margetan, F. J. and Thompson, R. B. and Yalda-Mooshabad, I.},
doi = {10.1007/BF00728250},
issn = {1573-4862},
journal = {Journal of Nondestructive Evaluation},
number = {3},
pages = {111--136},
title = {{Backscattered Microstructural Noise in Ultrasonic Toneburst Inspections}},
volume = {13},
year = {1994}
}

@article{Lan2015a,
author = {Lan, Bo and Lowe, Michael J S and Dunne, Fionn P E},
doi = {10.1016/j.jmps.2015.06.012},
issn = {0022-5096},
journal = {Journal of the Mechanics and Physics of Solids},
pages = {221--242},
publisher = {Elsevier},
title = {{A generalized spherical harmonic deconvolution to obtain texture of cubic materials from ultrasonic wave speed}},
url = {http://dx.doi.org/10.1016/j.jmps.2015.06.012},
volume = {83},
year = {2015}
}

@article{Lan2015c,
author = {Lan, Bo and Lowe, Michael J. S. and Dunne, Fionn P. E.},
doi = {10.1016/j.jmps.2015.06.014},
issn = {00225096},
journal = {Journal of the Mechanics and Physics of Solids},
pages = {179--198},
publisher = {Elsevier},
title = {{A spherical harmonic approach for the determination of HCP texture from ultrasound: A solution to the inverse problem}},
url = {http://dx.doi.org/10.1016/j.jmps.2015.06.014},
volume = {83},
year = {2015}
}

@article{Li2015c,
author = {Li, J. and Rokhlin, S. I.},
doi = {10.1016/j.wavemoti.2015.05.004},
issn = {01652125},
journal = {Wave Motion},
pages = {145--164},
title = {{Propagation and scattering of ultrasonic waves in polycrystals with arbitrary crystallite and macroscopic texture symmetries}},
volume = {58},
year = {2015}
}

@article{Ming2020,
author = {Huang, M. and Sha, G. and Huthwaite, P. and Rokhlin, S. I. and Lowe, M. J. S.},
doi = {10.1121/10.0002916},
issn = {0001-4966},
journal = {The Journal of the Acoustical Society of America},
number = {6},
pages = {3645--3662},
pmid = {33379920},
publisher = {Acoustical Society of America},
title = {{Elastic wave velocity dispersion in polycrystals with elongated grains: Theoretical and numerical analysis}},
volume = {148},
year = {2020}
}

@article{Rokhlin2021a,
author = {Rokhlin, S. I. and Sha, G. and Li, J. and Pilchak, A. L.},
doi = {10.1016/j.ultras.2021.106433},
issn = {0041624X},
journal = {Ultrasonics},
number = {March},
pmid = {34034095},
title = {{Inversion methodology for ultrasonic characterization of polycrystals with clusters of preferentially oriented grains}},
volume = {115},
year = {2021}
}

@article{Zhang2008,
author = {Zhang, Yonghong and Wang, Yu and Zuo, Ming J. and Wang, Xiaodong},
doi = {10.1109/CCECE.2008.4564854},
isbn = {9781424416431},
issn = {08407789},
journal = {Canadian Conference on Electrical and Computer Engineering},
pages = {1797--1800},
publisher = {IEEE},
title = {{Ultrasonic time-of-flight diffraction crack size identification based on cross-correlation}},
year = {2008}
}

@article{Yeoh2023,
   author = {W. Y. Yeoh and B. Lan and M. J. S. Lowe},
   doi = {10.1098/rspa.2023.0176},
   issn = {1364-5021},
   issue = {2277},
   journal = {Proceedings of the Royal Society A: Mathematical, Physical and Engineering Sciences},
   month = {9},
   title = {Investigation of the influence of macrozones in titanium alloys on the propagation and scattering of ultrasound},
   volume = {479},
   url = {https://royalsocietypublishing.org/doi/10.1098/rspa.2023.0176},
   year = {2023},
}

@article{Liu2021a,
   author = {Yuan Liu and Michał K. Kalkowski and Ming Huang and Michael J.S. Lowe and Vykintas Samaitis and Vaidotas Cicėnas and Andreas Schumm},
   doi = {10.1016/j.ultras.2021.106441},
   issn = {0041624X},
   issue = {March},
   journal = {Ultrasonics},
   pmid = {33894662},
   title = {Can ultrasound attenuation measurement be used to characterise grain statistics in castings?},
   volume = {115},
   year = {2021},
}

@article{Huang2020,
   author = {M. Huang and G. Sha and P. Huthwaite and S. I. Rokhlin and M. J. S. Lowe},
   doi = {10.1121/10.0002102},
   issn = {0001-4966},
   issue = {4},
   journal = {The Journal of the Acoustical Society of America},
   note = {},
   pages = {1890-1910},
   pmid = {33138527},
   publisher = {Acoustical Society of America},
   title = {Maximizing the accuracy of finite element simulation of elastic wave propagation in polycrystals},
   volume = {148},
   year = {2020},
}

@article{Huang2021a,
   author = {M. Huang and P. Huthwaite and S. I. Rokhlin and M. J. S. Lowe},
   doi = {10.1098/rspa.2021.0850},
   issn = {14712946},
   issue = {2258},
   journal = {Proceedings of the Royal Society A: Mathematical, Physical and Engineering Sciences},
   publisher = {Royal Society Publishing},
   title = {Finite-element and semi-analytical study of elastic wave propagation in strongly scattering polycrystals},
   volume = {478},
   year = {2022},
}

\end{document}